\documentclass[twocolumn]{aastex63}
\usepackage[utf8]{inputenc}
\pdfoutput=1 %for arXiv submission
\usepackage{amsmath,amstext}
\usepackage[T1]{fontenc}
\usepackage{lineno}
\newcommand{\htwo}        {\mbox{H$_{2}$}}
\newcommand{\jone}        {\mbox{$J=1-0$}}
\newcommand{\jtwo}        {\mbox{$J=2-1$}}
\newcommand{\jthree}        {\mbox{$J=3-2$}}
\newcommand{\jfour}        {\mbox{$J=4-3$}}
\newcommand{\jseven}        {\mbox{$J=7-6$}}

\newcommand{\kmpers}      {\mbox{\rm km~s$^{-1}$}}

\newcommand{\percmcu}     {\mbox{\rm cm$^{-3}$}}
\newcommand{\msun}        {\mbox{\rm M$_\odot$}}

\newcommand{\msunperyr}   {\mbox{\rm M$_\odot$~yr$^{-1}$}}

\newcommand{\xco}         {\mbox{$X_{\rm CO}$}}

\newcommand{\aco}         {\mbox{$\alpha_{\rm CO}$}}
\newcommand{\xcounits}    {\mbox{\rm cm$^{-2}$(K km s$^{-1}$)$^{-1}$}}
\newcommand{\acounits}  {\mbox{\rm M$_\odot$ (K km s$^{-1}$ pc$^2$)$^{-1}$}}

\newcommand{\percmsq}     {\mbox{cm$^{-2}$}}

\newcommand{\ann}[1]{{#1}}

\begin{document}

%\shortauthors{Bolatto et al.}

\author[0000-0002-5480-5686]{Alberto D. Bolatto}
\affiliation{Department of Astronomy, University of Maryland, College Park, MD 20742, USA}
\affiliation{Joint Space-Science Institute, University of Maryland, College Park, MD 20742, USA} 
\affiliation{Visiting Scholar at the Flatiron Institute, Center for Computational Astrophysics, NY 10010, USA}
\affiliation{Visiting Astronomer, National Radio Astronomy Observatory, VA 22903, USA}
\author[0000-0002-2545-1700]{Adam K. Leroy}
\affiliation{Department of Astronomy, The Ohio State University, Columbus, OH 43210, USA}
\author[0000-0003-2508-2586]{Rebecca C. Levy}
\affiliation{Department of Astronomy, University of Maryland, College Park, MD 20742, USA}
\author[0000-0001-9436-9471]{David S. Meier}
\affiliation{Department of Physics, New Mexico Institute of Mining and Technology, 801 Leroy Pl., Socorro, NM, 87801, USA}
\affiliation{National Radio Astronomy Observatory, P. O. Box O, 1003 Lopezville Rd., Socorro, NM, 87801, USA}
\author[0000-0001-8782-1992]{Elisabeth A. C. Mills}
\affiliation{Department of Physics and Astronomy, University of Kansas, 1251 Wescoe Hall Dr., Lawrence, KS 66045, USA}
\author[0000-0003-2377-9574]{Todd A. Thompson}
\affiliation{Department of Astronomy, The Ohio State University, Columbus, OH 43210, USA}
\author[0000-0001-6527-6954]{Kimberly L. Emig}
\altaffiliation{Jansky Fellow of the National Radio Astronomy Observatory}
\affiliation{National Radio Astronomy Observatory, 520 Edgemont Road, Charlottesville, VA 22903-2475, USA}
\author[0000-0002-3158-6820]{Sylvain Veilleux}
\affiliation{Department of Astronomy, University of Maryland, College Park, MD 20742, USA}
\affiliation{Joint Space-Science Institute, University of Maryland, College Park, MD 20742, USA} 
\author[0000-0001-8224-1956]{J\"urgen Ott}
\affiliation{National Radio Astronomy Observatory, P. O. Box O, 1003 Lopezville Rd., Socorro, NM, 87801, USA}
\author[0000-0001-9300-354X]{Mark Gorski}
\affiliation{Chalmers University of Technology, Department of Space, Earth, and Environment, SE-412 96 Gothenburg, Sweden}
\author[0000-0003-4793-7880]{Fabian Walter}
\affiliation{Max-Planck-Institut für Astronomie, Königstuhl 17, D-69120 Heidelberg, Germany}
\affiliation{National Radio Astronomy Observatory, P. O. Box O, 1003 Lopezville Rd., Socorro, NM, 87801, USA}
\author[0000-0002-1790-3148]{Laura A. Lopez}
\affiliation{Department of Astronomy, The Ohio State University, Columbus, OH 43210, USA}
\author[0000-0003-4023-8657]{Laura Lenki\'c}
\affiliation{Department of Astronomy, University of Maryland, College Park, MD 20742, USA}

\correspondingauthor{Alberto D. Bolatto}
\email{bolatto@umd.edu}

\begin{abstract}
We present the ALMA detection of molecular outflowing gas in the central regions of NGC\,4945, one of the nearest starbursts and also one of the nearest hosts of an active galactic nucleus (AGN). We detect four outflow plumes in CO \jthree\ at $\sim0.3\arcsec$ resolution that appear to correspond to molecular gas located near the edges of the known ionized outflow cone and its (unobserved) counterpart behind the disk. The fastest and brightest of these plumes has emission reaching observed line-of-sight projected velocities of over 450\,\kmpers\ beyond systemic, equivalent to an estimated physical outflow velocity $v\gtrsim600$\,\kmpers\ for the fastest emission. Most of these plumes have corresponding emission in HCN or HCO$^+$ \jfour. We discuss a kinematic model for the outflow emission where the molecular gas has the geometry of the ionized gas cone and shares the rotation velocity of the galaxy when ejected. We use this model to explain the velocities we observe, constrain the physical speed of the ejected material, and account for the fraction of outflowing gas that is not detected due to confusion with the galaxy disk. We estimate a total molecular mass outflow rate ${\rm \dot{M}_{mol}}\sim20$\,\msunperyr\ flowing through a surface within 100~pc of the disk midplane, likely driven by a combination of the central starburst and AGN.
\end{abstract}


\title{ALMA Imaging of a Galactic Molecular Outflow in NGC\,4945}

\shorttitle{A Galactic Molecular Outflow in NGC\,4945 }

\shortauthors{Bolatto, A. D., Leroy, A. K., Levy, R. C., et al.}

\section{Introduction}
\label{sec:intro}
NGC~4945, located in the southern hemisphere, is one of the nearest massive galaxies \citep[$3.8 \pm 0.3$~Mpc corresponding to a scale of 18.4\,pc per arcsecond;][]{Karachentsev2007}. The galaxy harbors a central starburst \citep{Bendo2016,Emig2020} as well as a Seyfert 2 Active Galactic Nucleus \citep[AGN,][]{Schurch2002}. Similar to the nearby starburst galaxies M\,82 and NGC~253, NGC~4945 is one of the brightest far-infrared sources outside the Local Group \citep{Sanders2003}. And like these two examples, it is one of the very few starbursts seen in gamma-ray emission by {\em Fermi}, although it is unclear if the emission is due to the star formation activity (and the ensuing supernovae) or to the central AGN \citep{Ackermann2012}.

NGC~4945 is viewed almost edge-on, with an axis ratio of nearly 5, corresponding to an inclination $i>80^\circ$. Inspection of the extinction apparent in optical and near-infrared images shows that the southwest rim of the galaxy lies closest to us. Thus the galaxy disk obscures much of what happens south of the nucleus while directions northwest of the nucleus are mostly free of extinction \citep[e.g.,][]{Marconi2000}. 

Partly because of this orientation, the AGN in NGC~4945 was discovered as a luminous but heavily absorbed ($\log[N_H/\percmsq]\sim24.7$) hard X-ray source with variability on time-scales of hours \citep{Iwasawa1993,Schurch2002,Marinucci2012}. In fact, NGC~4945 is the brightest Seyfert 2 galaxy in the $50-100$~keV sky \citep{Done1996}. Its black hole has a mass inferred from water maser measurements of $M_{BH}\sim1.4\times10^6$~\msun\ \citep{Greenhill1997}, similar to the central Milky Way black hole but accreting at a much larger rate. Nonetheless, analysis of mid-infrared spectroscopy suggests the AGN is not energetically dominant over the central region \citep{Forbes98,Spoon2000}. 

Imaging in the near-infrared also finds a $\sim100$~pc scale starburst ring in Paschen~$\alpha$, but no central point source corresponding to the AGN \citep{Marconi2000}. This starburst is likely fed by a bar, which is invisible due to the high inclination and extinction but inferred from kinematic modeling \citep{Lin2011}. The 93~GHz radio continuum emission from the central region is $\sim85\%$ thermal and implies a star formation rate of approximately $4.3$~\msun\,yr$^{-1}$, in agreement with far-infrared estimates \citep{Bendo2016}. A recent high-resolution study of the starburst by \citet{Emig2020} finds 27 compact ($1-4$\,pc), free-free dominated sources that likely correspond to young massive clusters powering the starburst. Of these, 15 are detected in H40$\alpha$ and H42$\alpha$ emission, and these radio recombination lines provide compelling evidence for the presence of young massive star clusters. These clusters produce $\gtrsim20-44\%$ of the total ionizing photon rate of the starburst. Meanwhile, the bright central compact source identified with the AGN is synchrotron-dominated and does not have associated radio-recombination emission. The limit to the mm-wave free-free emission from the central source implies that at least 90\% of its UV radiation is absorbed before before being able to ionize the surrounding gas \citep{Emig2020}. Therefore less than 10\% of the ionizing radiation from the AGN escapes into the nuclear starburst region.  

\begin{figure*}[t]
    \centering
    \includegraphics[width=\textwidth]{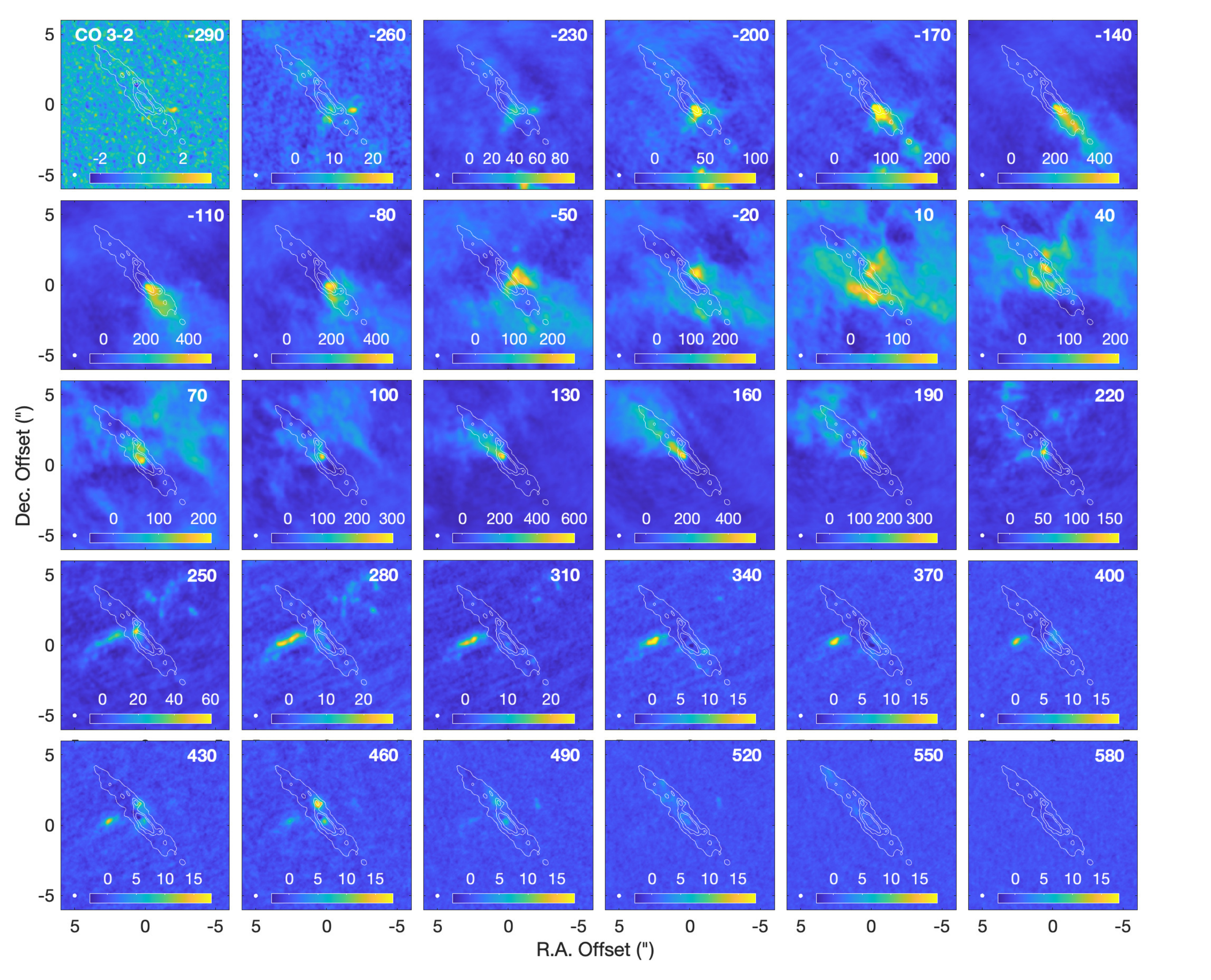}
    \caption{Channel maps for CO \jthree\ emission (linear scale $18.4$\,pc/\arcsec). The reference coordinate is $\alpha_{2000}=13^{\rm h}05^{\rm m}27\fs47$, $\delta_{2000}=-49^{\circ}28{\arcmin}05\farcs6$ for this and all other figures. Each panel shows the emission averaged over 30\,\kmpers, with the value in the upper right corner of each panel giving the velocity of that channel in \kmpers\ relative to the adopted systemic velocity of $v_{sys}=563$\,\kmpers. The white contours, which are the same in all panels, show the in-band continuum \ann{indicating the starburst region}, with contour values of  9.4, 25, and 37.5 mJy\,beam$^{-1}$. The color bar in each panel indicates the color stretch, with values in mJy\,beam$^{-1}$. The beam is $0.26\arcsec$, and the first two channels have higher noise because they primarily come from combining archival data obtained using a different (higher resolution) array configuration. The rotation of the galaxy disk is apparent at velocities between $v\sim-190$ and $190$\,\kmpers (\S\ref{sec:kinematics}), while emission at $v\lesssim-200$\,\kmpers\ and $v\gtrsim220$\,\kmpers\ corresponds mostly to the molecular outflow. \ann{The bright emission inside the contours around 460\,\kmpers\ is most likely not CO but neighboring transitions of H$^{13}$CN \jfour\ at rest frequencies of 345.34 GHz.}}
    \label{fig:channelsco}
\end{figure*}

\begin{figure}
%    \centering
    \includegraphics[width=1\columnwidth]{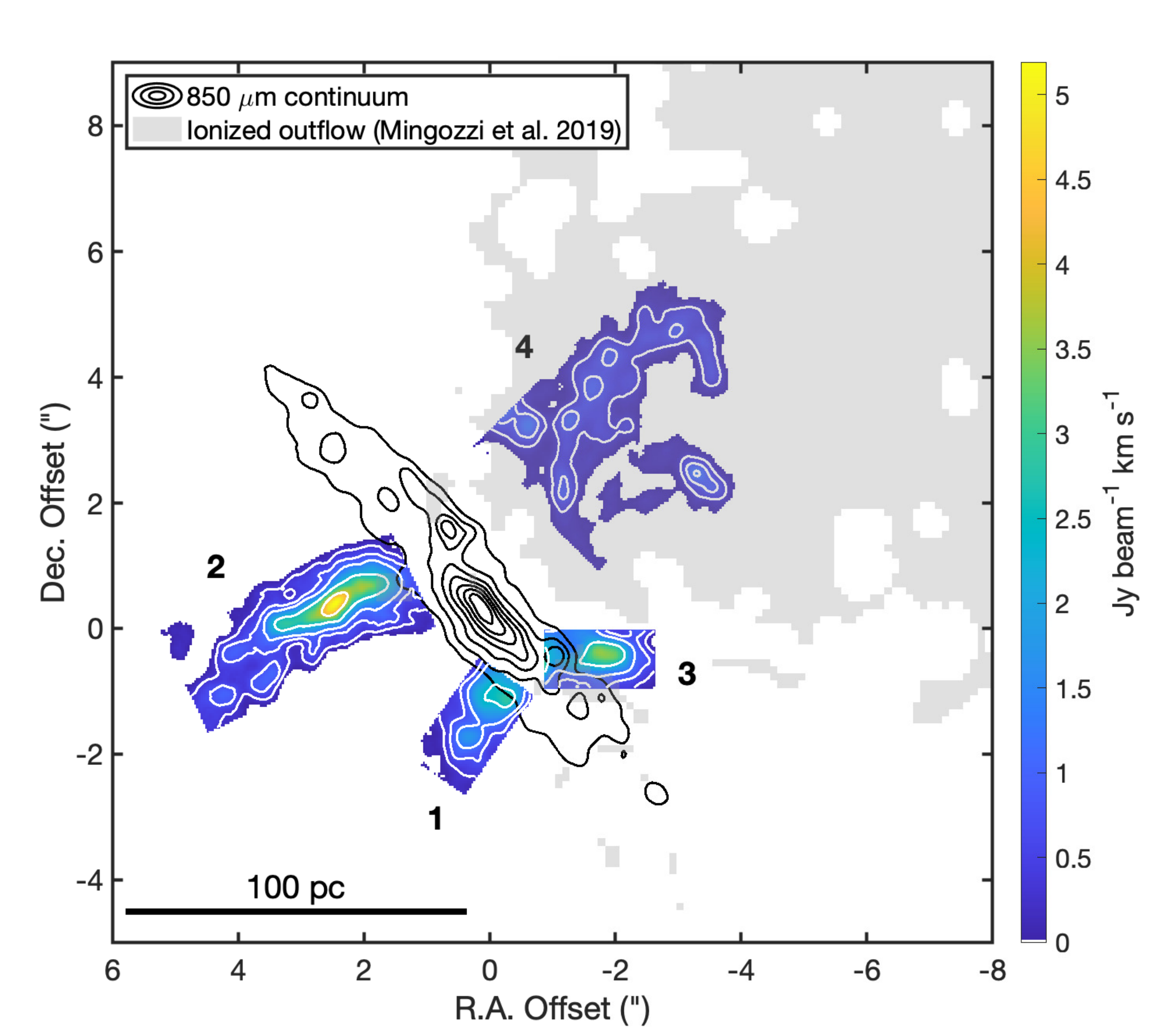}
    \caption{The molecular and ionized gas outflows in NGC~4945. The color scale shows the CO \jthree\ integrated intensity in the four plumes visible in the molecular outflow, numbered 1 to 4 here, with the spectra shown in Figure \ref{fig:regionspectra} integrated over their respective velocity ranges (Table \ref{tab:outflow}). The white contours inside the colored regions correspond to line-integrated CO intensities integrated over outflow velocities of 0.3, 0.7, 1, 1.2, 2.4, 4 Jy\,beam$^{-1}$\,\kmpers. The black contours illustrate the Band 7 continuum, starting at 10  mJy\,beam$^{-1}$ with contours stepping by 10 mJy\,beam$^{-1}$. The gray, transparent area corresponds to the optical emission identified by \citet{Mingozzi2019} as belonging to the outflow based on the BPT ionization classification. See Table \ref{tab:outflow} for more details on the individual outflow features.
    \label{fig:outflow}}
\end{figure}

\begin{figure}
    \centering
    \includegraphics[width=\columnwidth]{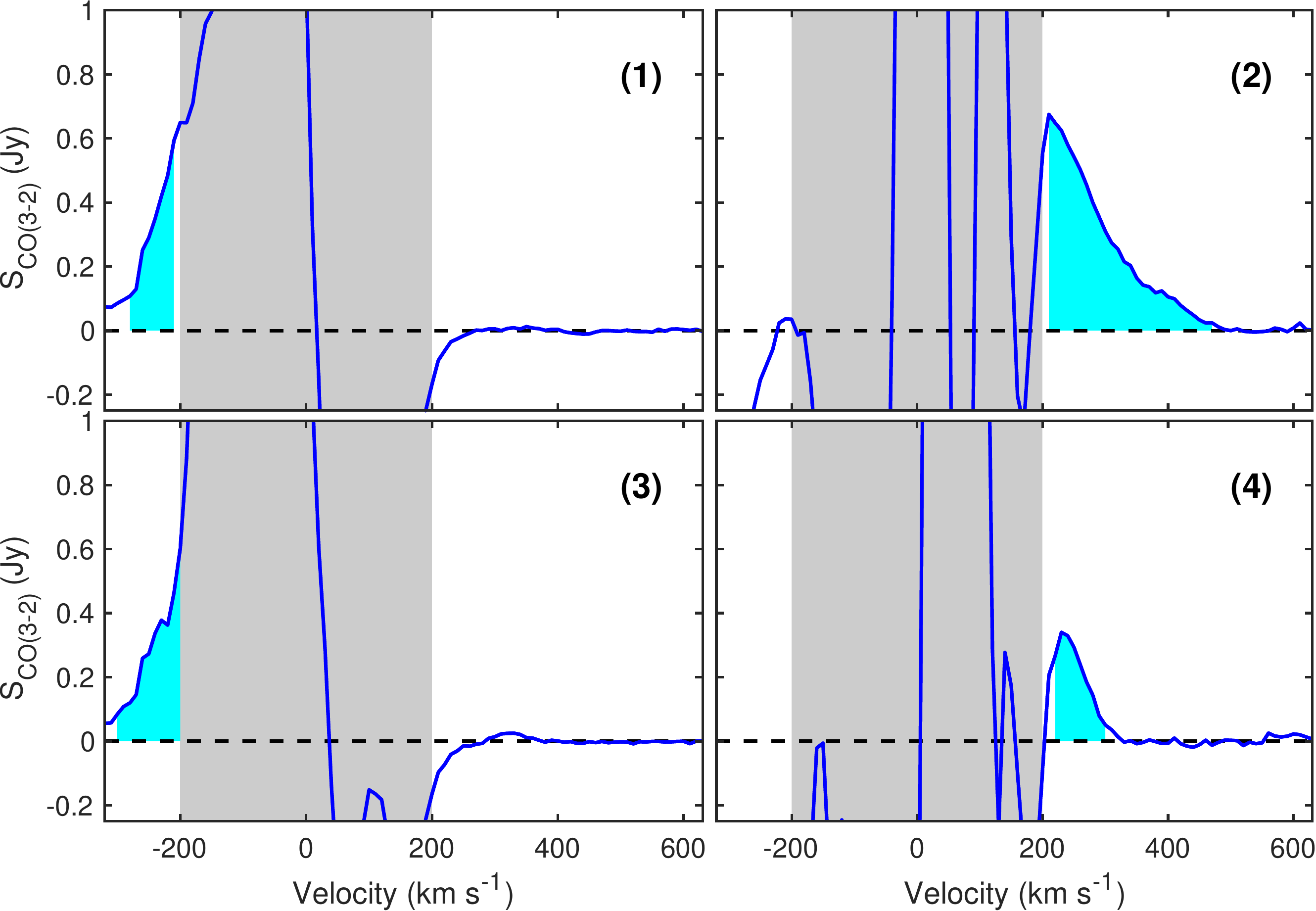}
    \caption{Spectra of the identified outflow regions. We show the area-integrated spectra of the four identified molecular outflow plumes numbered as (1) SW, (2) SE, (3) NW, and (4) NE. The velocities associated with the outflow emission are marked in cyan (Table \ref{tab:outflow}). The confused CO emission associated with the galaxy disk is present between $v\sim-200$ and $200$\,\kmpers, marked here with a gray band. The large negative excursions in the spectrum reflect ``bowling'' artifacts in the image, caused by incomplete sampling of the Fourier plane and problems associated with the deconvolution of complex structures. The molecular outflow features appear as tails that can reach velocities much larger than the rotation of the galaxy. This large velocity extent is particularly apparent for the brightest feature, shown in panel (2). \ann{As we discuss in \S\ref{sec:geometry}, the velocity configuration of the outflow features is naturally due to the outflowing gas sharing the rotation of the galaxy.} See Table \ref{tab:outflow} for more details on the individual outflow features.}
    \label{fig:regionspectra}
\end{figure}

NGC~4945 is also notable for hosting a nuclear outflow. The outflow was identified through the presence of optically-detected radial filaments emanating from the central region \citep{Nakai1989}, as well as H$\alpha$ spectroscopy that revealed ``split'' multi-component line profiles with velocities of $-250$ to $-550$ \kmpers\ relative to systemic 250~pc north of the nucleus \citep{Heckman1990}. Modeling of the spectral data suggests a conical flow with a (full) opening angle $\sim78^\circ$ and an axis oriented $57^\circ$ away from the line-of-sight \citep{Heckman1990}. Imaging in H$\alpha$ reveals a conical structure northwest of the nucleus with an opening angle of $\sim70^\circ$ and an orientation aligned with the minor axis of the galaxy \citep{Moorwood1996}, exactly the morphology expected for an ionized wind. The spectrum of the brightest emission reveals FWHM linewidths of 600 \kmpers. This cone corresponds to the approaching (i.e., tilted toward the observer) lobe of the outflow. The receding cone to the southeast is not apparent in these observations, which is to be expected because this part of the outflow would be located behind the extinction screen of the galaxy disk. 

Recent observations using the Multi Unit Spectroscopic Explorer (MUSE) integral field spectrograph show the full velocity structure of the ionized gas in the outflow \citep{Venturi2017,Mingozzi2019}. The observations show the presence of both blueshifted and redshifted emission in the approaching northwest lobe of the outflow, as well as the existence of part of the receding southeast lobe, 25\arcsec\ away from the nucleus and beyond the region most highly extinguished by the disk \citep{Mingozzi2019}. The morphology and kinematics of the H$\alpha$ emission are reasonably reproduced by a hollow cone oriented 75$^\circ$ away from the line of sight with an outer (full) opening angle of 70$^\circ$, an inner opening angle of 50$^\circ$ degrees, and a constant outflow velocity \citep{Venturi2017}, fairly consistent with \citet{Heckman1990} but in better agreement with the inclination of NGC\,4945. Because the opening angle is larger than the angle between the axis of the cone and the plane of the sky, both redshifted and blueshifted velocities are present in the visible outflow lobe. 

The optical ionized wind also has an X-ray counterpart. The X-ray wind is bright in the soft X-rays at $0.3-2$~keV. Chandra X-ray observatory imaging shows an edge-brightened approximately-conical plume extending 30\arcsec\ away from the nucleus \citep{Schurch2002}. The X-ray plume has a narrower opening angle than the H$\alpha$ emission, $40^\circ$ to $50^\circ$, and thus X-ray emitting gas appears to fill the hollow cone model derived from H$\alpha$. X-ray spectroscopy of the plume finds it arises from Solar abundance plasma at $T\sim10^7$~K \citep{Schurch2002}. The structure of the outflow is very similar to that observed in the X-rays in NGC~253 \citep{Strickland2000}, although the NGC~4945 outflow appears to be hotter and have a wider opening angle.

In this work we present Atacama Large Millimeter/submillimeter Array (ALMA) data on the cool molecular wind associated with the ionized gas. Cool winds may have a profound impact on the evolution of galaxies, because they have the potential to remove significant amounts of mass \citep{Veilleux2020}. \ann{\citet{Henkel2018} analyzed the molecular kinematics of the central regions at $\sim2\arcsec$ resolution, and interpreted position-velocity cuts along the minor axis taken at $\sim\pm4.2\arcsec$ from the center as resulting from a 100\,pc radius disk expanding at 50\,\kmpers\ with a component of inflow farther out due to a bar. Here we present clear evidence for a fast central polar molecular wind with geometry very similar to the observed ionized wind, contained inside the $\sim4\arcsec$ radius of the previous study.} To our knowledge, \ann{such a} molecular wind has not been previously discussed in the literature. We assume a distance to NGC\,4945 of 3.8~Mpc \citep{Karachentsev2007} and note that the galaxy lies in the Centaurus~A/M~83 group and is part of of the Centaurus~A subgroup \citep{Karachentsev2002}.

We describe the observations in \S\ref{sec:observations}, present and discuss the results in \S\ref{sec:results}, and summarize our conclusions in \S\ref{sec:conclusions}.

\section{Observations}
\label{sec:observations}

The main observations for this study come from project 2018.1.01236.S (PI: A. Leroy), which used ALMA to observe the nucleus of NGC~4945 in both Band 3 ($\nu \approx 88{-}102$~GHz) and Band 7 ($\nu \approx 343{-}357$~GHz). The Band 3 data have been presented by \citet{Emig2020}. Here we focus on the Band 7 data, which were obtained in configuration C43-3, with a representative resolution of $\theta\sim0.41\arcsec$ and a maximum recoverable scale of $\sim4.7\arcsec$ for CO. These observations use J1427-4206 as passband and flux calibrator and J1326-5256 as gain calibrator. We targeted four spectral windows centered at 344.523, 342.701, 354.565, and 356.378 GHz configured with bandwidths of 1.875 GHz and channel widths of 1.953 MHz (960 channels per window). The Band 7 observations were obtained on 08-Dec-2018 for a total of 2,740 seconds. Calibration was done using the {\tt CASA} version 5.6.2 pipeline. The data for {\tt NGC4945\_a\_07\_TM2} were downloaded from the archive, calibrated with the provided calibration script, and split into separate datasets corresponding to CO \jthree, HCO$^+$ \jfour, HCN \jfour, and CS \jseven. The tuning used is intended to observe all four transitions simultaneously, but it has the drawback that it results in limited velocity coverage of CO and HCN: it does not extend much blueward of the CO transition ($v\gtrsim-250$\,\kmpers) or redward of the HCN transition ($v\lesssim+200$\,\kmpers).

To supplement the frequency coverage of the CO \jthree\ line on the high frequency side, we also include Band 7 observations from project 2016.1.01135.S (PI: N. Nagar) obtained in configuration C40-5, which has a representative resolution $\theta\sim0.16\arcsec$ and a maximum recoverable scale of $1.9\arcsec$ for CO. These observations use the same passband, flux, and gain calibrators as the 2018.1.01236.S Band 7 observations. The correlator is configured in 4 spectral windows, two using low spectral resolution TDM mode for continuum and two using the 1.875 GHz mode with channel widths of 0.488 MHz and centered on 344.240 and 346.043 GHz. The observations were obtained on 06-Jul-2017 for a total of 2,253 seconds. The data for {\tt ngc4945\_a\_07\_TM1} were downloaded from the archive, calibrated with the provided calibration script, and the CO portion of the visibilities was split and incorporated into our imaging. 

Imaging was done using {\tt CASA} version 5.6.2, using the task {\tt tclean} with Briggs weighting and a robust parameter set to 0.5. We adopted $z=0.00188$ ($\sim564$~\kmpers) for the systemic redshift of all four molecular transitions imaged, and their frequencies were obtained from Splatalogue \citep{Remijan2016}, which adopts them from the Cologne Database for Molecular Spectroscopy \citep{Muller2005}. The HCN, HCO$^+$, and CS lines were imaged with 0.05\arcsec\ pixels. The CO data were imaged with 0.035\arcsec\ pixels to properly sample the convolving kernel we use to generate a round beam, as these data have a smaller and somewhat elongated beam due to the inclusion of the archival extended configuration data. After imaging, the CO data were convolved to a round $0.26\arcsec$ ($\sim 4.8$ pc) beam.  The synthesized beams for HCN, HCO$^+$, and CS are very similar ($\theta\approx0.40\arcsec\times0.35\arcsec$ or $\sim 7.4\times 6.4$ pc with PA$\approx-15^\circ$). All cubes were produced with 10\,\kmpers\ wide channels and cleaned with the multiscale algorithm, using scales 0, 6, and 22 pixels for CO and 0, 8, and 32 pixels for the other lines, with 10,000 iterations and a threshold of 0.5 mJy. In the rest of this manuscript, wherever we use relative coordinates they are referred to $\alpha_{2000}=13^{\rm h}05^{\rm m}27\fs47$ and $\delta_{2000}=-49^{\circ}28{\arcmin}05\farcs6$, the phase center for the 2018.1.01236.S observations.

The resulting RMS ($1\sigma$) sensitivity in 10\,\kmpers\ channels is $0.9$~mJy\,beam$^{-1}$ for CS ($\nu\sim342.24$~GHz), $0.6$~mJy\,beam$^{-1}$ for CO ($\nu\sim345.15$~GHz, much worse at $v\lesssim-230$\,\kmpers\ where only the C40-5 data contributes), $1.7$~mJy\,beam$^{-1}$ for HCN ($\nu\sim353.84$~GHz), and $1.5$~mJy\,beam$^{-1}$ for HCO$^+$ ($\nu\sim356.06$~GHz). Note that these are approximate sensitivities for ``line free'' channels, while the images are heavily dynamic-range limited and have in practice considerably worse sensitivity wherever bright emission is present. \ann{The effective RMS sensitivity in CO in a channel at moderate velocities ($\sim100$\,\kmpers) with bright emission is closer to $\sim20$~mJy, resulting in an effective dynamic range of $\sim50$ driven by imperfect imaging of the bright extended emission.}

\section{Results and Discussion}
\label{sec:results}

\begin{deluxetable*}{rccccccccc}
%\tablewidth{0pt}
\tablecaption{Molecular Outflow Features \label{tab:outflow}}
\tablehead{
\colhead{Feature} & \colhead{Location} &  \colhead{Velocity range} & \colhead{I$_{\rm CO}$(\jthree)} & \colhead{M$_{\rm mol}$} & \colhead{$v_{disk}$} & \colhead{$v_{outf}^{proj}$} & \colhead{$d_{outf}^{proj}$} & \colhead{Apparent $\dot{\rm M}_{outf}$} & \colhead{Corrected $\dot{\rm M}_{outf}$}\\
& & \colhead{(\kmpers)} & \colhead{(Jy\,\kmpers)} & \colhead{($10^5$\,\msun)} & \colhead{(\kmpers)} & \colhead{(\kmpers)} & \colhead{(pc)} &  \colhead{(\msun\,yr$^{-1}$)} & \colhead{(\msun\,yr$^{-1}$)}
}
\startdata
1 & SW & $[-280,-210]$ & $26.0$ & $2.2$ & $-160$ & $-75$  & $21$ & $0.8$ & $3.6$\\
2 & SE & $[+210,+460]$ & $69.5$ & $5.9$ & $+160$ & $+117$ & $34$ & $2.0$ & $2.0$\\
3 & NW & $[-300,-200]$ & $34.9$ & $2.9$ & $-160$ & $-80$  & $12$ & $2.0$ & $3.0$\\
4 & NE & $[+220,+300]$ & $26.6$ & $2.2$ & $+160$ & $+83$ & $70$ & $0.3$ & $1.4$
\enddata
\tablecomments{Properties of the four features associated with the molecular outflow. See spectra in Figure \ref{fig:regionspectra} and spatial locations in Figure \ref{fig:outflow}. The two southern features (SW and SE) are in the receding outflow cone, and the two northern features (NW and NE) are in the approaching outflow cone. The velocity range indicates the velocities used for the ${\rm I_{CO}}$ integration of the spectra in Figure \ref{fig:regionspectra}. The molecular mass in the features, ${\rm M_{mol}}$, is computed using Equation \ref{eq:mmol}. The value $v_{disk}$ indicates the rotation velocity from the nearby disk emission that was subtracted from the first moment of the region spectrum to obtain $v^{proj}_{outf}$, the line-of-sight intensity-weighted projected velocity of the outflowing gas. The value $d^{proj}_{outf}$ indicates the projected distance between the spatial barycenter of the outflow feature and the major axis of the nuclear region. The values for apparent $\dot{\rm M}$ and corrected $\dot{\rm M}$ indicate the molecular mass loss rate measured using Equation \ref{eq:mdot} for the projected quantities with $\tan i_\varphi=1$, and corrected for projection effects (mostly due to velocity, \S\ref{sec:corrected}) respectively.}
\end{deluxetable*}

\begin{figure}
    \centering
    \includegraphics[width=\columnwidth]{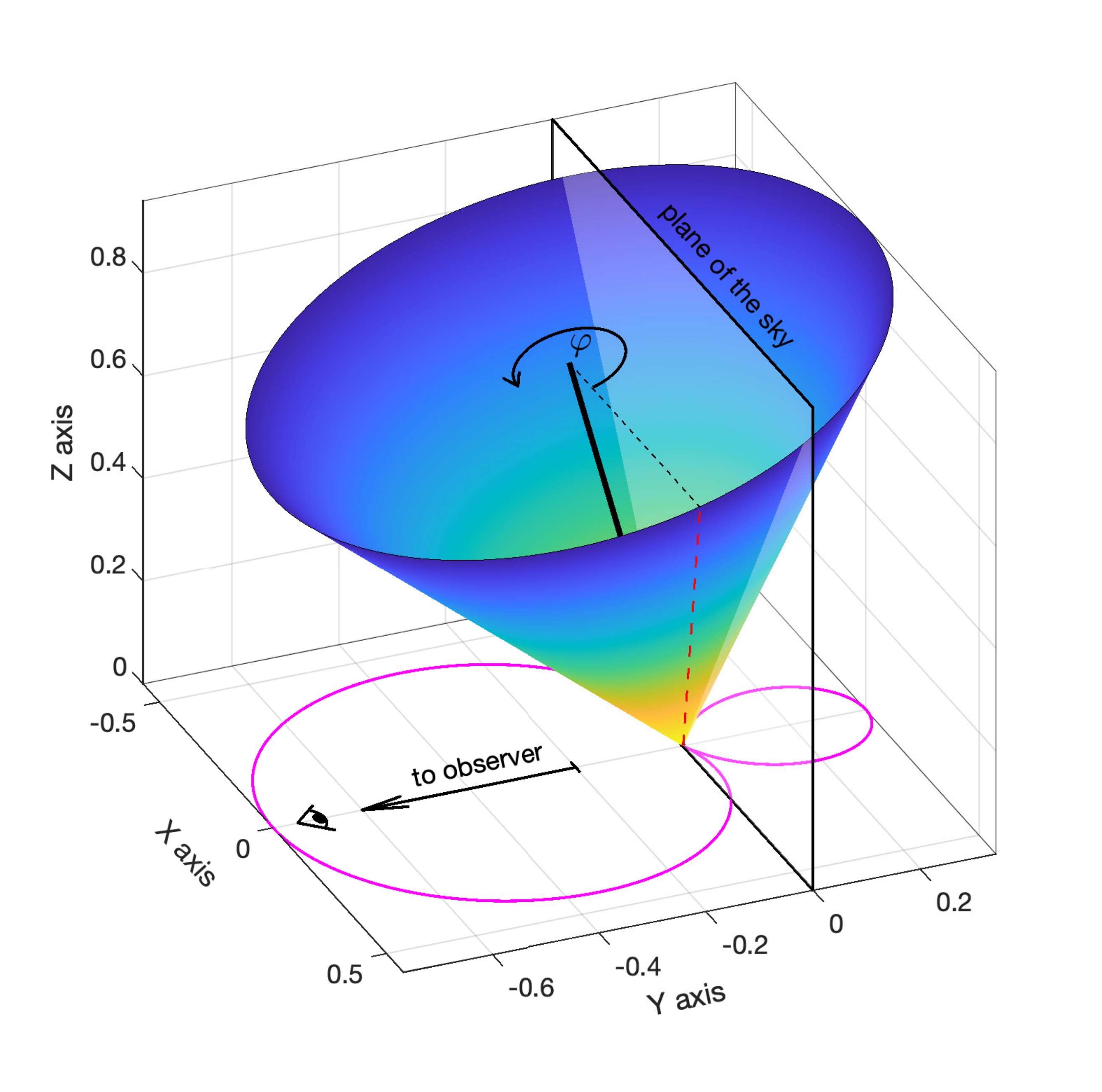}
    \caption{A view of the geometry of the approaching side of the outflow. The cone has an axis with inclination of 75$^\circ$ with respect to the line of sight (the $y$ axis), a semi-opening angle $\alpha=35^\circ$, and the ruled surface has unity length. The plane of the sky is shown at $y=0$, intersecting the cone. The magenta line in the XY plane shows the projection factor of the outflow velocity on the line-of-sight, $\cos i_\varphi$, with the polar plot presenting the cosine of the inclination angle projected along the line-of-sight to the observer $i_\varphi$ as a function of azimuthal angle $\varphi$. The corresponding minimum velocity deprojection factor is $1.3$. The line of $\varphi=0$ is shown dashed-red on the surface of the cone, and corresponds to the intersection of the cone with the $y=0$ plane for positive $x$ before rotating it toward the observer by the angle $90^\circ-i$. The observer is located at $\varphi=270^\circ$ in this system.}
    \label{fig:outflowcone}
\end{figure}

\subsection{The Molecular Wind in CO Emission}

Inspection of the channel maps of the CO \jthree\ observations clearly shows molecular features that are not part of the rotation pattern of the central regions (Figure \ref{fig:channelsco}). Note that because the tuning of the 2018.1.01236.S observations was designed to simultaneously obtain CO~\jthree\ and HCN~\jfour\ (as well as HCO$^+$~\jfour\ and CS~\jseven), velocities below $-240$~\kmpers\ for CO and over $+200$~\kmpers\ for HCN fall outside the passband. Nonetheless, after merging with the archival observations for 2016.1.01135.S which use a different tuning, we were able to recover information for lower CO velocities. These archival observations, however, were obtained in a more extended configuration, which yields higher angular resolution but results in much lower sensitivity to extended structure for CO velocities $v<-240$~\kmpers. 

The channels at $v=-260$ and $-230$ \kmpers\ in Figure \ref{fig:channelsco} show a V-like emission pattern arising from the SE of the starburst region delineated by the contours corresponding to the dust continuum emission. The emission shows two plumes launched symmetrically at an angle of $\sim60^\circ$ from the line of the major axis of the galaxy, about 0.6\arcsec\ SE of the center.

Emission from the rotating disk of the galaxy first becomes visible between $v=-200$ and  $-170$~\kmpers and continues until $v\sim+200$~\kmpers . The CO emission in the central channels, from about $-140$~\kmpers\ to $+160$~\kmpers , is very highly confused in our data. In this range the emission reflects the combined effects of the high optical depth of the line, self-absorption, possible absorption against the dust continuum, and missing spatial frequencies due to the incomplete sampling of the $u{-}v$ plane by the employed interferometer configurations. As a result, while the outflow may contribute emission to these channels, distinguishing features associated with outflowing gas in the range of $v\sim-200$ to $+200$~\kmpers\ is almost impossible. 

At velocities above $v\sim+200$~\kmpers\ two features incompatible with a rotating disk appear: a prominent narrow plume of emission south of the nucleus, pointing toward the east, that is reminiscent of the SW streamer emerging from the NGC~253 nucleus \citep{Bolatto2013a}. The second feature is a collection of clumps and diffuse emission located northwest of the nucleus that is prominent around $v=+250$ to $+280$~\kmpers. The CO spectra of individual clumps show long velocity tails stretching to $v\sim+300$\,\kmpers, unlike normal disk clouds.

Note that the features at $-230$~\kmpers as well as the prominent plume at $\sim+280$~\kmpers\ appear to have their ``foot points'' at the southwest and northeast edges of the brightest 850~$\mu$m continuum. This region is associated with the dustiest part of the nuclear starburst and the AGN and corresponds to sources 17, 18, and 20 in \citet{Emig2020}. Sources 17 and 20 are candidate young clusters. The central source, source 18, is likely a combination of a nuclear cluster and the AGN.

The features that we observe are characteristic of a galactic molecular wind, which must be driven by the activity in the central regions of NGC~4945. Ionized gas features associated with a wind are visible farther away from the highly extincted central regions and have been previously characterized as discussed in \S1 \citep{Nakai1989,Heckman1990,Moorwood1996, Mingozzi2019}.

We number features associated with the outflow in CO emission $1-4$ following the order SW, SE, NW, and NE (Figure \ref{fig:outflow}). The SW feature is associated with the receding outflow cone at negative velocity; the SE feature is associated with the receding outflow cone at positive velocity; the NW feature with the approaching outflow cone at negative velocity; and the NE feature is associated with the approaching outflow cone at positive velocity.

Spatially-integrated spectra extracted over a rectangular region encompassing each feature are shown in Figure \ref{fig:regionspectra}, with the velocities that we identify as outflow marked in cyan, with the corresponding integrated intensity map in Figure \ref{fig:outflow}. The brightest feature, the SE feature that we label number 2, presents emission well beyond the rotation velocity of the galaxy. The range of velocities covered by this feature is large, and the emission decays steadily away from $+200$\,\kmpers\ but it is detectable all the way up to $+480$\,\kmpers. The velocity ranges and integrated fluxes adopted for each feature are listed in Table \ref{tab:outflow}. There we also report the line-of-sight mean velocity in excess of the apparent rotation velocity of the disk at the base of each feature.  We estimate the reference rotation velocity to be $v_{disk}=-160$\,\kmpers\ for features in the west (1 and 3) and $v_{disk}=+160$\,\kmpers\ for features in the east (2 and 4), by looking for the closest emission to the foot points that we can clearly associate with disk kinematics. This is consistent with our derived inner rotation curve (\S\ref{sec:kinematics}).

\begin{figure}
    \centering
    \includegraphics[width=\columnwidth]{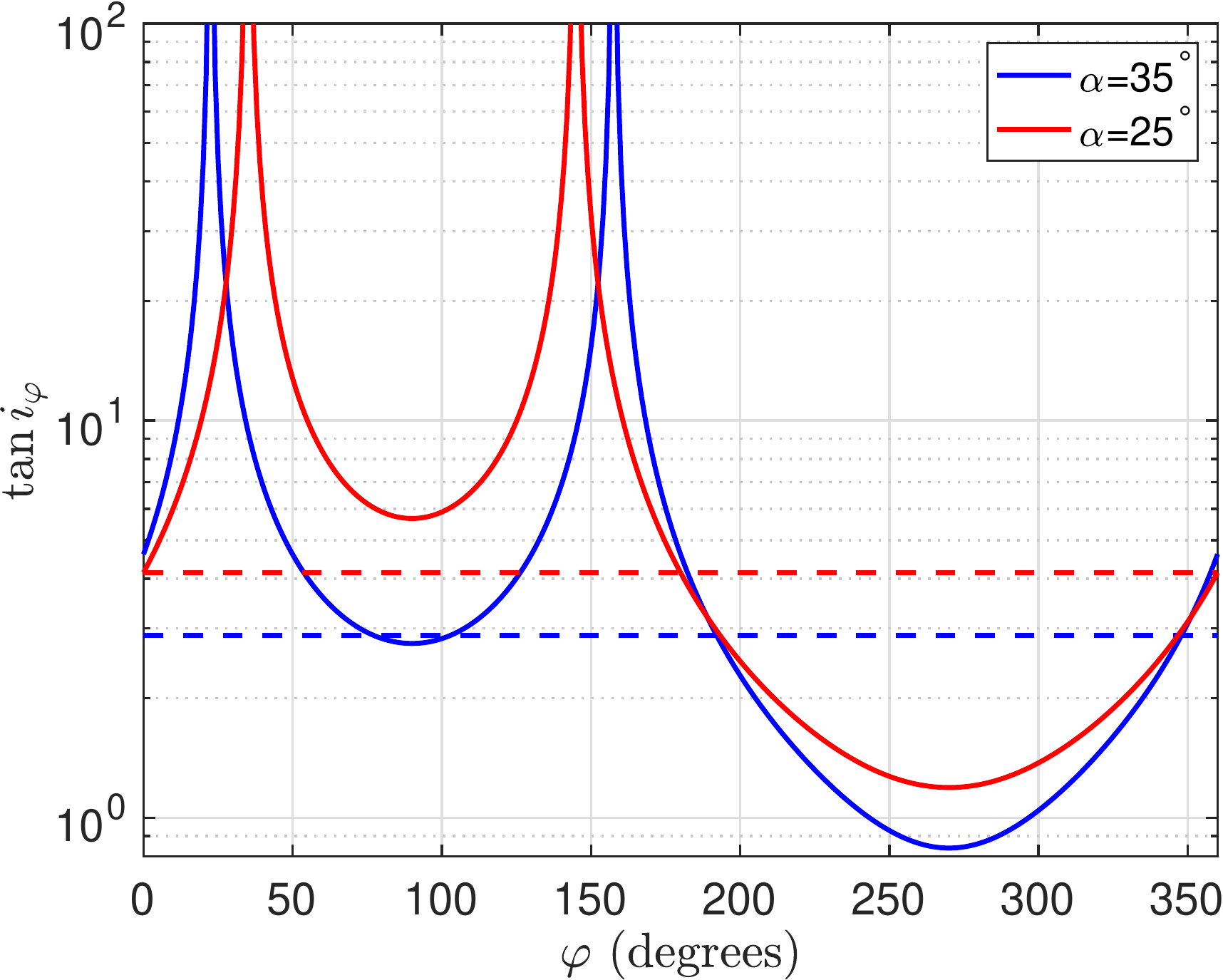}
    \caption{Geometric deprojection factor $\tan i_\varphi$ as a function of azimuthal angle, $\varphi$, around the cone described in Figure~\ref{fig:outflowcone} for two values of the cone semi-opening angle, $\alpha$. The median value of the deprojection factors of 2.9 and 4.1 for $\alpha=35^\circ$ and $25^\circ$ respectively, are indicated by the dashed lines.}
    \label{fig:deprojfactor}
\end{figure}

\subsection{Apparent Molecular Mass of the Outflow}
\label{sec:outflowrate}

We compute the mass outflow rate leaving the nuclear region using the measured CO \jthree\ line fluxes, the projected velocities, and length of each wind feature reported in Table \ref{tab:outflow}. Given a known distance and an estimate of the CO-to-H$_2$ conversion factor, the molecular mass associated with a CO line-integrated flux in the \jone\ transition can be computed using Equation (3) in \citet{Bolatto2013b}. To use our observed \jthree\ flux density we convert it to a \jone\ flux density using the equation

\begin{equation}
    \frac{S_{\rm CO}(J_m\rightarrow J_{m-1})}{S_{\rm CO}(J_l\rightarrow J_{l-1})}=\left(\frac{J_m}{J_l}\right)^2\,r_{JmJl},
\end{equation}

\noindent where $J_m=3$, $J_l=1$, and $r_{31}$ is the line ratio in Rayleigh Jeans brightness temperature units (K). A ratio $r_{31}=1$ corresponds to the high density ``thermalized'' case at high temperature, where the level populations follow a Boltzmann distribution. In general in optically thick emission $r_{31}$ will be lower than unity, reflecting the fact that the gas is less excited than in the high-density, high-temperature limit. For example, the value of $r_{31}$ in pointings in the outflow of M\,82 is on average $r_{31}\sim0.5-0.6$ \citep{Weiss2005,Leroy2015}. In galaxy disks this ratio increases with indicators of star formation activity \citep{Lamperti2020}, with typical line ratios in the disks of lower excitation ``normal'' galaxies of $r_{31}\sim0.2-0.4$ (Leroy et al. ApJ submitted). For our computation we assume an intermediate excitation $r_{31}=0.5$, resulting in $S_{\rm CO}(\jone)=S_{\rm CO}(\jthree)/4.5$. Moreover, the CO-to-\htwo\ conversion value in starbursting galaxy centers is frequently observed to be lower than the observed value in massive giant molecular clouds in the Milky Way galactic disk \citep{Bolatto2013b}. Following the large-velocity-gradient and dust-based calculations of \citet{Leroy2015} for the M\,82 outflow and the considerations of \citet{Zschaechner2018} for the NGC\,253 outflow, we adopt $\xco=0.5\times10^{20}$\,\xcounits\ \ann{(equivalent to $\aco=1.1$\,\acounits)} for the NGC\,4945 outflow, a quarter of the typical value for the Galactic disk. With these considerations the molecular mass associated with the observed CO emission, in \msun, becomes

\begin{equation}
    {\rm M_{\rm mol}}=583\,S_{\rm CO}(\jthree)\,\Delta\,v\,D_{\rm Mpc}^2,
    \label{eq:mmol}
\end{equation}

\noindent with $S_{\rm CO} \Delta\,v$ in Jy\,\kmpers, and $D_{\rm Mpc}=3.8$, our assumed distance. The masses reported in Table \ref{tab:outflow} employ this equation, which contains the mass contribution of He. \ann{Note that what matters to determining the mass is the ratio of the assumed \xco\ and $r_{31}$, and we are assuming a conservatively low \xco\ as it corresponds to warm gas. If we were to increase the excitation to $r_{31}\sim0.8-1.0$ (the maximum for optically thick emission from warm gas), the inferred M$_{\rm mol}$ would drop by a factor of $1.6-2$. This is likely a lower boundary to the effect of excitation and conversion factor on M$_{\rm mol}$, since the emission we observe is very unlikely to be optically thin. If, by contrast, we were to decrease $r_{31}$ to typical ``normal galaxy'' values or increase the conversion factor to Milky Way disk molecular cloud values M$_{\rm mol}$ would increase by a factor of $1.6-4$.} 

To compute the mass outflow rate associated with the molecular phase we use \citep[e.g.,][]{Bolatto2013a}

\begin{equation}
    \dot{\rm M}_{outf}= 10^{-6}\frac{{\rm M_{mol}}\,|v_{outf}^{proj}|}{d_{outf}^{proj}}\tan{i_\varphi}
    \label{eq:mdot}
\end{equation}

\noindent where $v_{outf}^{proj}$ and $d_{outf}^{proj}$ are the mean line-of-sight velocity and on-the-sky distance traveled by the ejected gas, in \kmpers\ and parsecs respectively, and the prefactor is used to convert from \kmpers\ to pc\,yr$^{-1}$. Figure \ref{fig:outflowcone} illustrates the geometry we use to describe the outflow cone pointing toward the observer (i.e., approaching). The axis of the cone is inclined by an angle $i_a$ with respect to the line of sight, and all outflow features are assumed to be radial following a generatrix of the conical surface identified by its azimuthal angle $\varphi$. The tangent of the inclination of a generatrix with respect to the line-of-sight, $i_\varphi$, accounts for the effect of deprojecting both the observed velocity (which is corrected by $(\cos i_\varphi)^{-1}$) and distance traveled (which is corrected by $(\sin i_\varphi)^{-1}$), and it is in general a function of the azimuthal angle $\varphi$. For a cone with semi-opening angle $\alpha$ the line-of-sight inclination $i_\varphi$ takes values in the range $i_a+\alpha\geq i_\varphi \geq i_a-\alpha$, with the extremes at $\varphi=90^\circ$ and $\varphi=270^\circ$ respectively in our defined coordinate system. The apparent $\dot{\rm M}_{outf}$ for each of our outflow features is listed in Table \ref{tab:outflow} in \msun\,yr$^{-1}$, computed for $\tan{i_\varphi}=1$.

For galaxies that are close to edge-on, one of the largest uncertainties in the computation of mass outflow rates is the geometry of the outflowing gas. Indeed, the $\tan{i_\varphi}$ factor becomes arbitrarily large as the inclination approaches $90^\circ$, driven by the deprojection of the measured line-of-sight velocity. The inclination of the central regions of the galaxy appears to be close to $i\approx75^\circ$ \citep{Henkel2018}, which would result in a ``typical'' deprojection factor of $\tan{i}\sim3.7$. Our own measurement of the HCN \jfour\ and CS \jseven\ kinematics (\S\ref{sec:kinematics}) is consistent with $i\approx73^\circ\pm3^\circ$. As discussed in \S\ref{sec:intro}, the optical measurements of the ionized wind are consistent with a cone of $i_a\approx75^\circ$ (normal to the galaxy plane), with semi-opening angle between $\alpha=35^\circ$ and $25^\circ$ \citep{Mingozzi2019}. If we assume the geometry of the molecular wind cone is described by $i_a=75^\circ$ and $\alpha=35^\circ$, it is possible for the inclination angle of a feature in the outflow to be as small as $i_\varphi=40^\circ$ for $\varphi=270^\circ$ (when it points directly toward the observer), yielding $\tan i_\varphi\approx0.84$. This is a very unlikely geometry as we will see in \S\ref{sec:geometry}, which means that in all likelihood $\tan i_\varphi>1$ and the apparent mass loss in Table \ref{tab:outflow} is a lower limit. Also note that the only phase considered here is the molecular phase, while galactic outflows are clearly multi-phase. 

\begin{figure*}
    \centering
    \includegraphics[width=\textwidth]{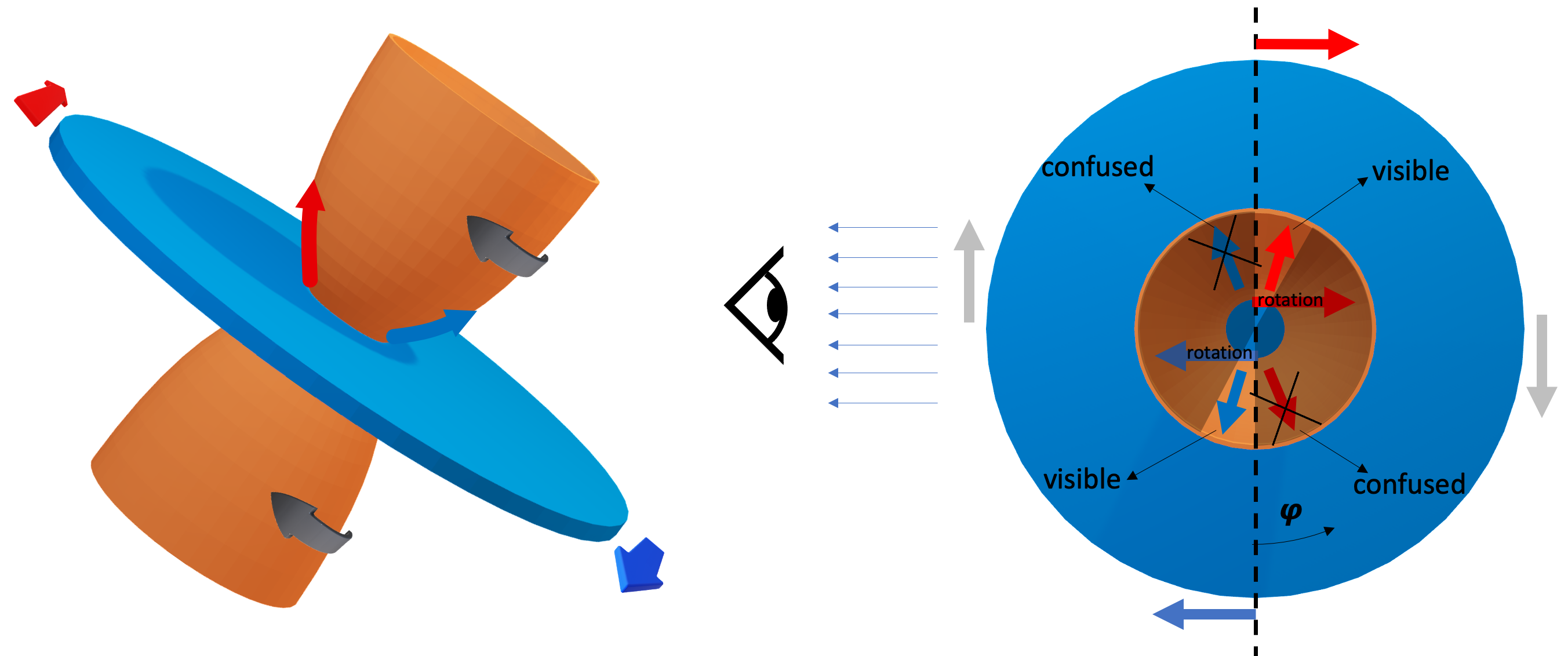}
    \caption{Velocities and velocity confusion in a highly inclined outflow. The left side shows a inclined view of an idealized galaxy outflow with a configuration similar to NGC\,4945, seen from the viewpoint of the observer. The blue disk represents the midplane of the galaxy, while the orange truncated hollow paraboloids show the outflow lobes. The disk rotates, with the right side approaching and the left side receding. The gas ejected in the outflow has a mostly radial motion shown by the curved arrows on the edges of the approaching lobe (which approximately correspond to features 3 and 4 in NGC\,4945), but close to the galaxy disk it also preserves the angular momentum and rotation motion of the original disk material. The rotation of the lobes is indicated by the gray arrows. On the right side we show the top view of an outflow with $i=90^\circ$ (orthogonal to the observer) for simplicity. The dashed line represents the plane of the sky, and the horizontal and vertical arrows the projection of the rotation velocity with their color representing the sign of the projection along the line-of-sight. We also indicate the azimuthal angle $\varphi$. The radial blue and red arrows inside the hollow orange paraboloid represent the radial motion of the outflow gas colored by the sign its projection on the line-of-sight. Regions of the outflow where the projection of the radial outflowing motion adds to the rotation have line-of-sight velocities in excess of the rotation velocity of the galaxy disk and are therefore distinct when seen in projection (see \S\ref{sec:geometry}, this is the ``velocity separation'' condition). Regions where the projection of the radial outflowing motion subtracts from the rotation have line-of-sight velocities lower than the rotation of the disk are confused with emission from the disk. The darkened sectors (with radial arrows crossed out) show the conceptual regions where confusion occurs and the outflowing gas may not be identified as such. The precise azimuthal wedge over which outflowing material is separated in velocity and thus not confused depends on the outflow velocity, the rotation velocity, the inclination, and the opening angle, but we will preferentially detect approaching outflowing gas on the blue side and receding outflowing gas on the red side of the disk (just like in NGC\,4945).}
    \label{fig:geometry}
\end{figure*}

\subsection{Geometry of the Molecular Wind}
\label{sec:geometry}

Figure \ref{fig:outflowcone} shows the idealized rendition of the approaching outflow lobe, which corresponds to the lobe containing CO features 3 and 4 and the visible ionized gas emission (Figure \ref{fig:outflow}). Molecular gas is usually thought to be in a layer surrounding the ionized gas, which in turn is in a hollow cone surrounding the very hot X-ray emitting outflow \citep[e.g.,][]{Leroy2015}. This is validated by the close association between CO emission in resolved outflows and H$\alpha$ filaments \citep[e.g.,][]{Bolatto2013a}. Physical simulations also suggest a close connection between H$\alpha$ and H$_2$ at the interface of the hot and cold phases, caused by turbulent mixing \citep{Fielding2020}. We adopt the parameters inferred from the ionized gas observations, $i_a=75^\circ$ and the outer $\alpha=35^\circ$, to describe the geometry of the molecular outflow cone. This does not account for the thickness of a molecular/neutral layer, but given the uncertainties it is a sufficiently good tractable description. The surface of the cone is traced by the unity vector $\hat{r}$ (the generatrix) as it revolves around the axis of the cone, $\hat{a}$, with a constant semi-opening angle $\alpha$. The azimuthal angle for the rotation of $\hat{r}$ around $\hat{a}$ is $\varphi$. The cone axis is inclined $i_a$ with respect to the line of sight $-\hat{y}$, such that $\cos i_a=-\hat{a}\cdot\hat{y}$. The magenta line shows the polar plot corresponding to the projection of the $\hat{r}$ vector along the line of sight, $\cos i_\varphi=-\hat{r}\cdot\hat{y}$, as a function of the azimuthal angle $\varphi$. As expected, for gas moving along a generatrix the maximum velocity projection toward the observer ($\varphi=270^\circ$) corresponds to $|\cos(75^\circ-35^\circ)|=0.766$, and the maximum velocity projection away from the observer ($\varphi=90^\circ$) is $|\cos(75^\circ+35^\circ)|=0.342$.

The geometric deprojection factor in Eq. \ref{eq:mdot} for different portions of the outflow corresponds to the tangent of their inclination, $\tan i_\varphi$, and it has a large effect on the mass outflow rate for portions of the outflow that are close to the plane of the sky due to the velocity deprojection. Figure \ref{fig:deprojfactor} shows this factor as a function of azimuthal angle $\varphi$ for the nominal inclination $i_a$ and two values of the semi-opening angle of the outflow cone, $\alpha$ \citep{Mingozzi2019}. The deprojection factor is larger than unity over 85\% (100\%) of the possible orientations for $\alpha=35^\circ$ ($\alpha=25^\circ$), and the median value of ${\tan i_\varphi}$ (the correction factor applicable to a fully populated cone) is $2.9$ ($4.1$). It is clear, however, that the appropriate value of the deprojection factor to infer the mass loss rate depends strongly on the azimuth of the outflowing gas. 

Can we further constrain the location of the outflowing gas on the outflow cone? One of the striking features of our outflow detection is that gas at both negative and positive velocities is present in both the approaching and receding outflow cones. Moreover, negative outflow velocities are seen in the outflow gas on the side of the galaxy where the disk is rotating toward us, while positive outflow velocities are on the side of the galaxy rotating away from us, irrespective of whether we are looking at the approaching or the receding outflow cone. Some aspects of this situation are similar to those present in the outflow in NGC\,253, which is also highly inclined \citep{Bolatto2013a,Krieger2019}. 

In order to better understand how this configuration arises, we can reason through a simple toy model representing an outflow in an edge-on galaxy (Figure \ref{fig:geometry}). The basic assumption is that the outflowing gas shares the rotation of the galaxy disk from which it originates. This is supported by observations of CO in M\,82, which find that the outflowing gas shares the angular momentum of the galaxy with slower rotation at greater distances from the midplane \citep{Seaquist2001,Walter2002,Leroy2015}. For outflowing gas to be easily identified as such it needs to be kinematically distinct from the gas in the midplane of the galaxy against which it is projected.  As we can see in Figure \ref{fig:channelsco}, that is most easily done for emission that has $v>|v_{rot}|$, where $v_{rot}$ is the rotation velocity of the galaxy at the galactocentric radius of interest. Indeed, for channels with $v\leq|v_{rot}|$ the CO emission from possible outflowing gas is completely confused with the background and foreground of widespread disk emission.
That means that against the approaching side of the disk we can most easily see outflow gas with velocities that are bluer than the rotation of the disk. That gas is in the part of the outflow pointing toward the observer, with a radial velocity that adds to the projected rotation. A similar situation occurs on the receding side of the disk. Therefore preferentially we will identify red outflowing emission on the receding side of the disk, and blue outflowing emission on the approaching side of the rotating disk (Figure \ref{fig:geometry}). 

\begin{figure}
    \centering
    \includegraphics[width=\columnwidth]{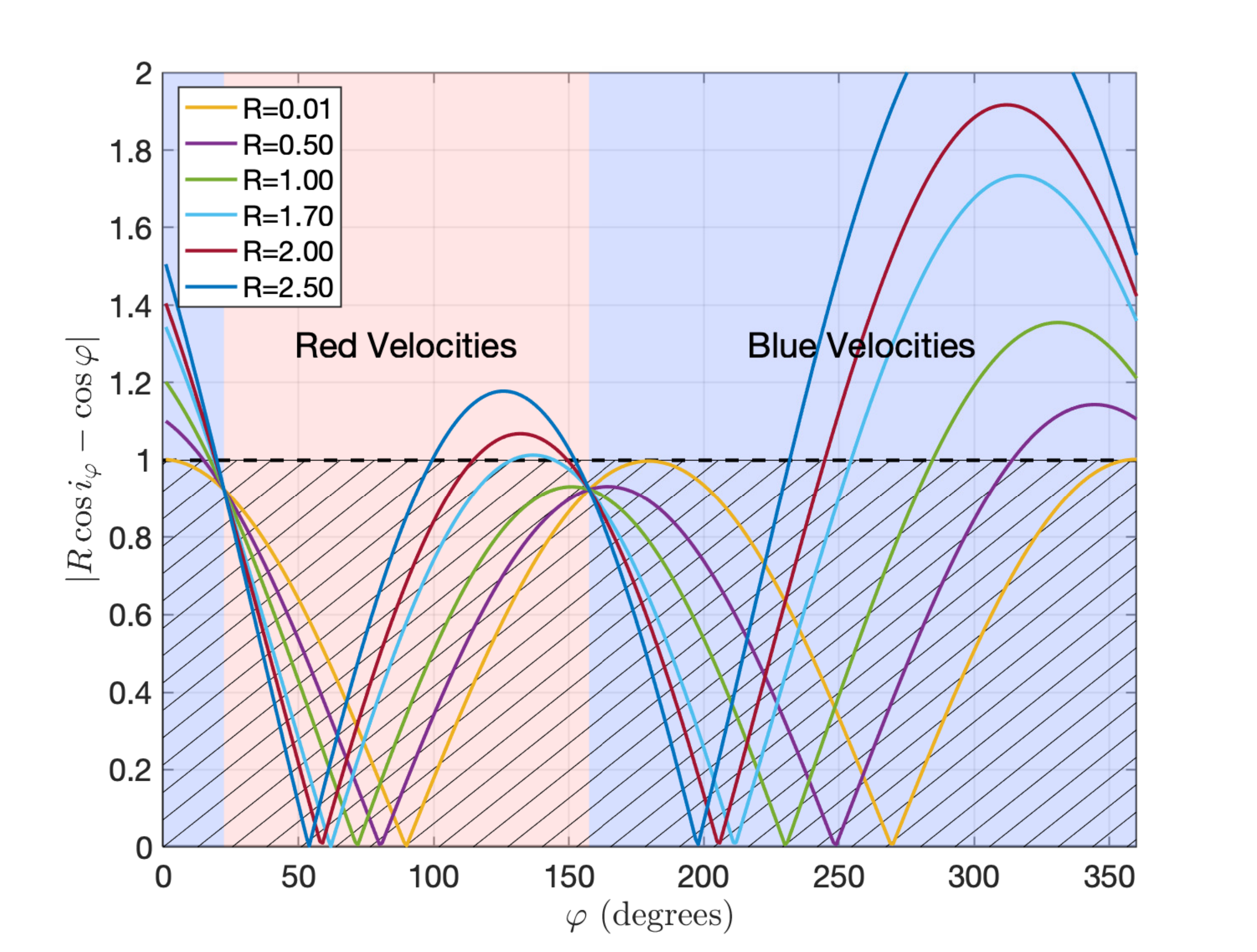}
    \caption{Velocity separation condition (Eq. \ref{eq:velsepNorth}) for the lobe of the NGC\,4945 outflow pointing towards us (Northern) plotted against the azimuthal angle in the cone, $\varphi$. Only regions where the curves are over unity fulfill the condition and are in principle visible (non-hatched region). The red shading indicates the region of positive (receding) velocities corresponding to the part of the cone that is behind the plane of the sky in Fig. \ref{fig:outflowcone}, the blue shading corresponds to the region of negative (approaching) velocities of the cone. The lines show results for different values of the ratio $R=v_{outf}/v_{rot}\sin{i}$. For $1.7<R<2.5$ it is possible to see outflowing gas at red velocities in the northern outflow cone, at azimuth angles of $100^\circ<\varphi<150^\circ$.}
    \label{fig:outflowangle}
\end{figure}

\begin{figure*}
    \centering
    \includegraphics[width=\textwidth]{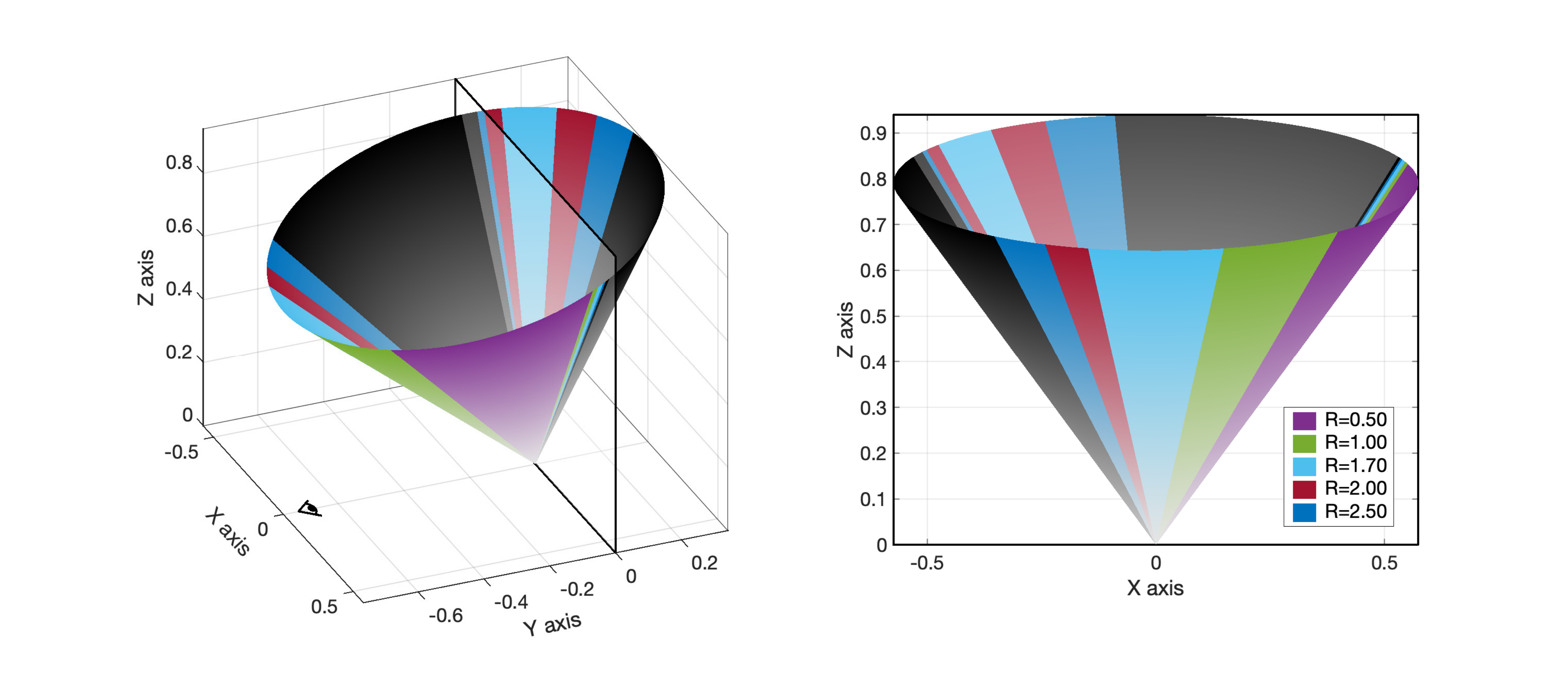}
    \caption{Rendering of the possible angle ranges in $\varphi$ for different values of the $R$ parameter: a representation of the angle ranges that fulfill the velocity separation condition in Figure~\ref{fig:outflowangle} presented on the surface of the cone. The left side shows the full view of the outflow cone including the plane of the sky at $y=0$. The right side shows the view from the observer viewpoint; as in Figure \ref{fig:outflowcone} the observer is located on $y=-\infty$. The color patches correspond to the $\varphi$ ranges that fulfill the velocity separation condition for five of the $R$ values shown in Figure~\ref{fig:outflowangle}, with ranges excluded shown in gray. Lower $R$ values, which result in more restricted $\varphi$ ranges that overlap with those for higher $R$ values, are shown on top: therefore the green range corresponding to $R=1.00$ also includes the range shown in purple for $R=0.50$. Note that the slower outflow velocities result in narrower $\varphi$ ranges that tend to outline the edges of the cone as seen by the observer. Also, there is a pileup of azimuthal angles at the edges of the cone: the purple range, which covers $\Delta\varphi\approx62^\circ$ (over one sixth of the cone surface), appears as a narrow sliver to the observer. The combination of these two projection effects naturally results in most CO emission approximately outlining the outflow cone.
    \label{fig:viewoutflow}}
\end{figure*}

The requirement that the emission from the outflow is not confused with gas in the disk experiencing the rotation of the galaxy can be expressed as a ``velocity separation'' condition 

\begin{equation}
    |v_{outf}\cos{i_\varphi}\pm v_{rot}\cos{\varphi}\sin{i_a}| > v_{rot}\sin{i_a},
    \label{eq:velsep}
\end{equation}

\noindent where the $+$ sign applies when the rotation of the galaxy is in the same sense as $\varphi$ (counterclockwise looking down along the outflow axis), and the $-$ sign when it is opposite. For NGC\,4945 the applicable sign is negative, the configuration we present in the first panel of Figure \ref{fig:geometry}. This equation becomes a condition on the possible azimuthal angles $\varphi$ for which outflowing gas can be easily identified given some ratio $R$ between the outflow velocity and the projected maximum rotation velocity, $R=v_{outf}/v_{rot}\sin{i}$, with the assumption that the axis of the outflow cone is normal to the disk of the galaxy, so $i_a=i$. For the north outflow cone in NGC\,4945 (which is pointing toward us), we can refactor the ``velocity separation'' condition as 

\begin{equation}
    |R\cos{i_\varphi}-\cos{\varphi}|>1
    \label{eq:velsepNorth}
\end{equation}

\noindent which we show in Figure \ref{fig:outflowangle} for several values of the ratio $R$ and for the adopted cone geometry of $i_a=75^\circ$ and $\alpha=35^\circ$. The range of $\varphi$ for which a stream in the outflow points away from the observer, thus resulting in red outflow velocities, is shaded light red ($22.5^\circ<\varphi<157.5^\circ$). The rest of the azimuthal angles result in outflow streams with velocities toward the observer, and are shaded light blue. Figure \ref{fig:viewoutflow} is a visualization of the velocity separation condition on the surface of the north outflow lobe in NGC\,4945. The left panel shows the full 3D view including the $y=0$ plane of the sky, the right panel shows the configuration from the observer's viewpoint. For simplicity we only present five of the $R$ values computed for Figure \ref{fig:outflowangle}, and use the same color-coding as in that Figure. Only $R\geq1.70$ are large enough ratios for outflow streams to have a red-shifted component (on the left side of the cone, as seen by the observer) with velocities beyond rotation. These same values of $R$ have the potential to produce blue-shifted outflow streams over a wide range of $\varphi$, as the range for $R=1.70$ seen in light-blue (for example) includes also the ranges for $R=1.00$ and $R=0.50$ in green and purple respectively.

Note that the velocity separation requirement is fulfilled for any value of $R$ at blue velocities given the favorable orientation of the cone. To easily detect an outflowing stream with red velocities on the lobe pointing toward us, however, requires gas with outflowing speeds at least 1.7 times faster than the rotation. That gas would be located at $\varphi\sim100^\circ-150^\circ$, thus have a deprojection factor $\tan{i_\varphi}\gtrsim3$ (Figure \ref{fig:deprojfactor}). In other words, detection of red velocities in the northern lobe of the NGC\,4945 outflow strongly suggests large corrections for any line-of-sight velocity and apparent mass loss rate measured. Region 4 in the outflow shows a general correspondence to this geometry, so it appears likely that its actual outflow speed is at least 3 times larger than the $v_{outf}^{proj}$ listed in Table \ref{tab:outflow}, yielding $v_{outf}\gtrsim240$\,\kmpers\ and maybe considerably higher. The mathematics are symmetric for the lobe of the outflow pointing away from the observer. The outflow region with a corresponding geometry in the southern lobe is region 1, which is also very likely to have a large correction to its projected outflow velocity.

Note also that CO streams with low outflow velocities, consequently low $R$, preferentially appear close to the edge of the cone, as seen in projection by the observer (Figure \ref{fig:viewoutflow}). Perhaps more importantly, the edges of the cone encompass a wide range of $\varphi$: if outflow gas were uniformly distributed in $\varphi$, we would see most of it toward the cone edges. A clear example is the $R=0.50$ purple wedge in Figure \ref{fig:viewoutflow}, which despite spanning over a sixth of the circumference projects as a narrow sliver on the western edge of the approaching lobe. This suggests a natural explanation as to why usually these structures are ``limb brightened'', with bright CO emission near the apparent edge of the ionized gas outflow cone \citep[as is the case in NGC\,253,][]{Bolatto2013a}.  

A consequence of these geometrical considerations is that there is a wide range of azimuthal angles for the outflow over which emission from outflowing gas will be confused with that from the rotating disk. Therefore, in outflows from highly inclined galaxies, we most easily detect emission over a wedge of azimuthal angles (notionally, those not shaded on the right side of Figure \ref{fig:geometry}, or those not in gray in Figure \ref{fig:viewoutflow}) and may miss a significant amount of gas with unfavorable projected kinematics. These considerations explain why in highly inclined galaxies the outflow emission in molecular gas is frequently found on narrow streams on the edge of the outflow cone. Moreover, for this reason our accounting of the outflowing gas in Table \ref{tab:outflow} is likely to be woefully incomplete: over a large range of azimuthal angles it is hard to separate the emission of the outflowing cone from the background of the galaxy disk. Note also that the outflow cone may not be fully populated with streams, and that not all streams may have the same outflowing velocities. In general, however, the easiest place to find streams with the lowest outflowing velocities is near the edges of the cone as seen by the observer. This is particularly true for gas on the approaching side of the galaxy rotation for the cone pointing toward the observer, and for gas on the receding side of the galaxy rotation for the cone pointing away from us.

Note that the velocity separation condition is a practical requirement, not an absolute one. With good quality spectral imaging and ancillary geometric information it is possible to model the kinematics and identify gas that is moving at velocities incompatible with rotation even if the projected velocity is smaller than the rotation of the disk. \ann{What requiring velocity separation does is to make it easy to identify outflowing material,} especially when the imaging is complex as it is for interferometric observations of widespread emission from a high optical depth transition.

\subsection{True Mass Outflow Rate}
\label{sec:corrected}

Taking into account these considerations it seems very likely that the apparent mass loss rate values we calculate in Table~\ref{tab:outflow} are lower limits to the mass loss for the features that have red outflow velocities in the approaching (N) cone and blue outflow velocities in the receding (S) cone, that is features 1 and 4. Interestingly, those are the features with the smallest apparent outflow rates. Figure~\ref{fig:outflowangle} shows the most likely azimuthal angle for red features in the approaching cone is $\varphi\sim130^\circ$. The corresponding deprojection factor for a semi-opening angle $\alpha=35^\circ$ is $\tan{i_\varphi}\simeq4.5$, overwhelmingly due to velocity deprojection (Figure \ref{fig:deprojfactor}). This means the corresponding corrected mass-loss rate for features 1 and 4 are $\dot{\rm M}_{mol}\sim3.6$ and 1.4 \msun\,yr$^{-1}$ respectively, with actual (intensity weighted) ejection velocities $v_{outf}\sim340$ and 380~\kmpers. Note that given the observed rotation velocity of $v_{rot}\sin{i}\simeq155$\,\kmpers\ (\S\ref{sec:kinematics}) the resulting ratio $R$ is $R\sim2.3$, consistent with the requirement from the ``velocity separation'' condition for emission to be visible. Note also that as $\varphi$ approaches that corresponding to the plane of the sky ($\varphi\simeq158.5^\circ$) the deprojection factor becomes arbitrarily large, while in the other direction the minimum value of $\tan i_\varphi$ possible is close to the median deprojection factor of 2.9. Therefore the corrected mass-rates can be at most 35\% smaller than the ones we adopt here, but they could as well be much larger.  

For features with blue velocities on the approaching cone (feature 3), or with red velocities on the receding cone (feature 2), the ``velocity separation'' condition does not constrain particularly well the azimuthal angle. The large velocity tail of feature 2 extending to over 450~\kmpers\ from systemic (Figure \ref{fig:regionspectra}) suggests that the deprojection correction could be small for this feature. Note that even with the smallest possible velocity deprojection allowed by our geometry, $1/\cos{40^\circ}$, this implies the physical velocity of the gas reaches out to $\sim600$~\kmpers\ from systemic. Morphologically, however, it seems unlikely that this feature is at the center of the outflow cone pointing directly toward us ($\varphi\sim270^\circ$) and much more likely that is close to the projected edge ($\varphi\sim310^\circ-20^\circ$). This suggests we should apply some mild boosting factor to the apparent velocity and mass-loss rate due to projection effects of order $\sim2$ (Figure \ref{fig:deprojfactor}). To be conservative and given the lack of constraints, we will instead assume $\tan{i_\varphi}\sim1$ for feature 2 and simply use the apparent mass outflow rate. If feature 2 were precisely at the minimum boost corresponding to $\varphi=270^\circ$ the mass-loss rate value should be reduced by 16\%, but we could also be underestimating it by 300\% and more. The real extent of feature 3 in projected velocity, on the other hand, is unclear because of the limitations of our tuning. Nonetheless, the symmetry with feature 1 and its placement with respect to the visible ionized gas cone also suggest that it is not emerging around $\varphi\sim270^\circ$ but much more toward the edge of the approaching cone, which would result in at least some mild boosting. If we conservatively assume $\varphi\gtrsim325^\circ$ based purely on the feature being very close to the edge of the ionized cone seen in projection (Figure \ref{fig:outflow}), the resulting boosting factor is $\tan{i_\varphi}\gtrsim1.5$. Therefore the corrected mass loss rate is $\dot{\rm M}_{mol}\gtrsim3.0$~\msun\,yr$^{-1}$, and the corrected, intensity weighted outflow velocity is $v_{outf}\gtrsim100$~\kmpers.

The sum total of the corrected molecular mass outflow rate for all four features detected in the outflow is therefore $\dot{\rm M}_{mol}\approx10$~\msun\,yr$^{-1}$. We have argued, however, that because of the geometry and the confusion with the disk emission we are limited to detecting only the portion of the outflowing gas for which geometry is favorable. According to the calculations presented in \S\ref{sec:geometry}, only a fraction of all possible azimuthal angles around the cone fulfill the ``velocity separation'' condition, and that fraction is a function of $R$, the ratio of outflow velocity to the projected maximum rotation velocity. For $R=1.7$, 2, and 2.5 the fraction of azimuthal angles $\varphi$ that fulfill Eq. \ref{eq:velsepNorth} are approximately $0.39$, $0.47$, and $0.56$ of the full circumference respectively. In other words, given the observed rotation velocity $v_{rot}\approx155$~\kmpers\ (\S\ref{sec:kinematics}), for outflow velocities of up to $v_{outf}\sim400$~\kmpers\ we would expect to detect as outflow only $\sim50\%$ of all possible directions in the outflow cone. 

\begin{figure*}[t]
    \centering
    \includegraphics[width=\textwidth]{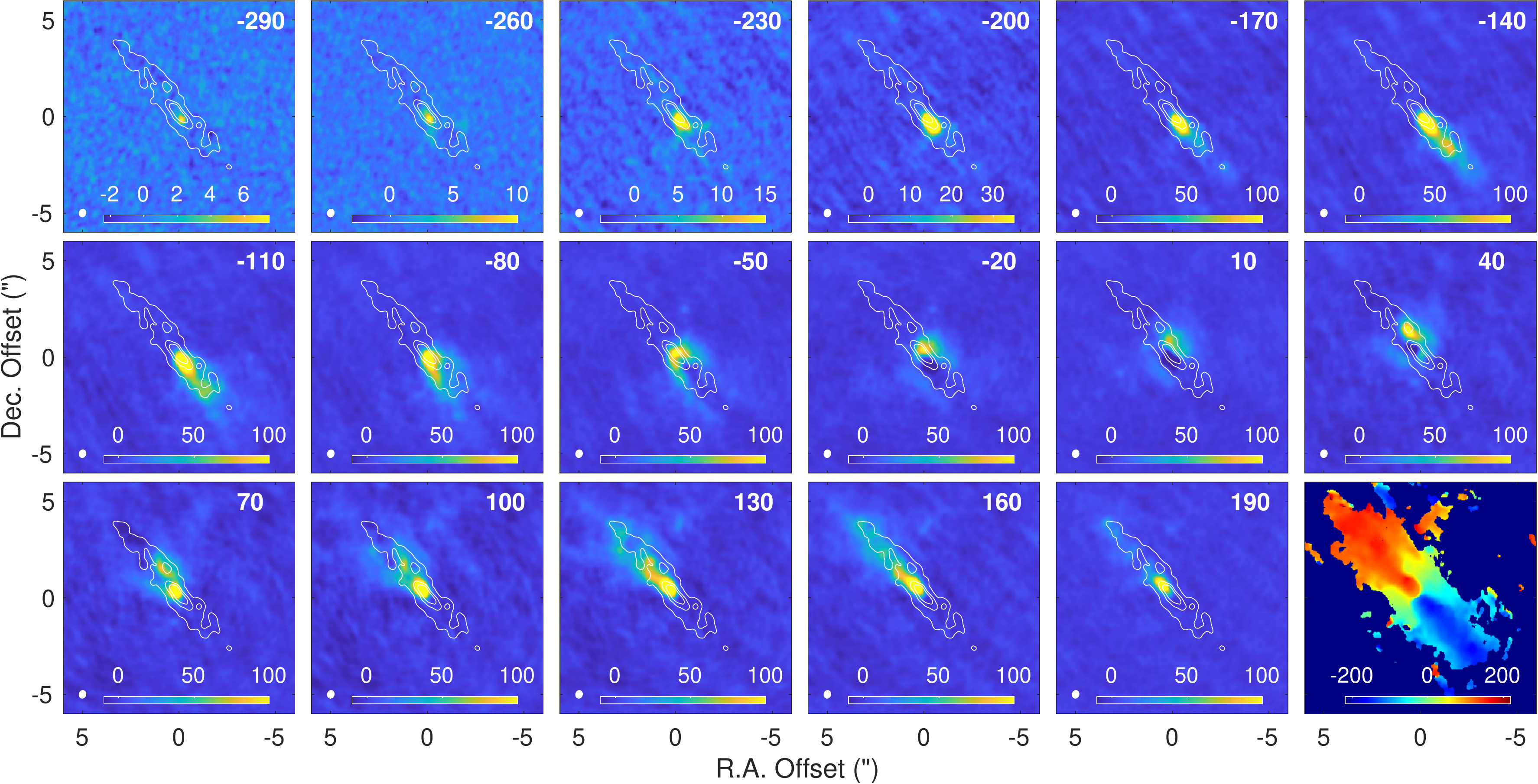}
    \caption{Channel maps for HCN \jfour\ emission. As in Figure \ref{fig:channelsco} each panel shows the emission averaged over 30\,\kmpers, with velocities referred to the systemic velocity ($v_{sys}=563$\,\kmpers) noted on the top right corner (in \kmpers). The white contours show the in-band (846 $\mu$m) continuum at 15, 25, and 37.5 mJy\,beam$^{-1}$. The color bar in each panel indicates the color stretch, with values in mJy\,beam$^{-1}$. The beam is $0.40\arcsec\times0.34\arcsec$ with PA=$-15^\circ$ (white ellipse on the bottom left corner of each panel). The last panel corresponds to the velocity field used to derive the harmonic decomposition in Figure \ref{fig:rotation}, obtained by Gaussian fitting. The HCN is at the red end of the USB passband to simultaneously observe CO in the LSB, therefore the highest receding velocity observed is $v=+190$\,\kmpers, missing the brightest feature (feature 2) seen in CO. Faint emission corresponding to features 1 and 3 is present at $v=-230$\,\kmpers. \ann{The faint emission at offsets (-0.2\arcsec, -3.75\arcsec) and $100-160$\,\kmpers\ could be part of the disk, or possibly an example of the confused approaching molecular outflow cone since it is located at the edge of the ionized outflow in projection.}
    \label{fig:channelshcn}}
\end{figure*}

\begin{figure*}[t]
    \centering
    \includegraphics[width=\textwidth]{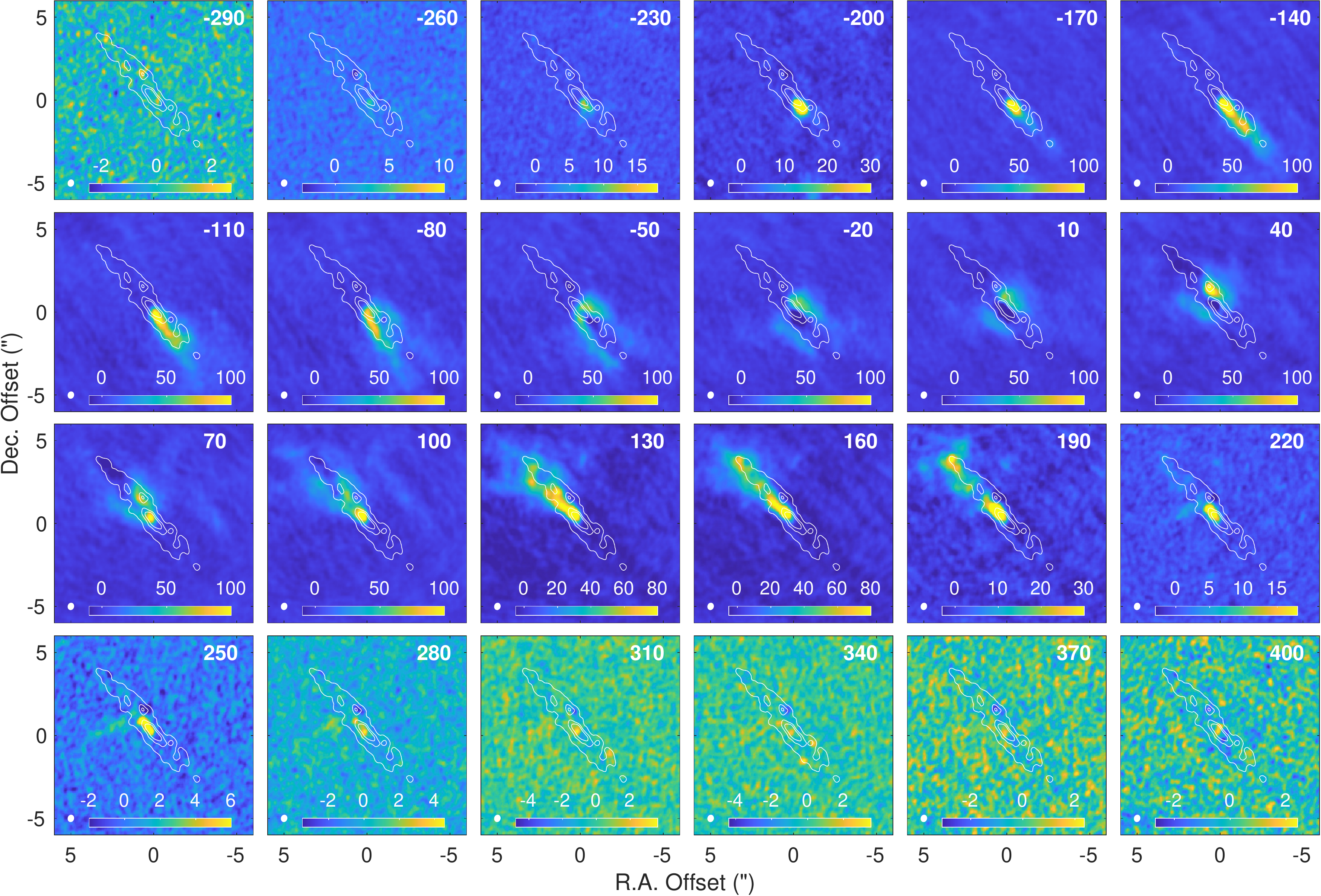}
    \caption{Channel maps for HCO$^+$ \jfour\ emission. Contours, color bars, and beam are as in Figure \ref{fig:channelshcn}. The emission corresponding to outflow feature 2 is apparent at $v=220-250$\,\kmpers.  
    \label{fig:channelshcop}}
\end{figure*}

\begin{figure*}[t]
    \centering
    \includegraphics[width=\textwidth]{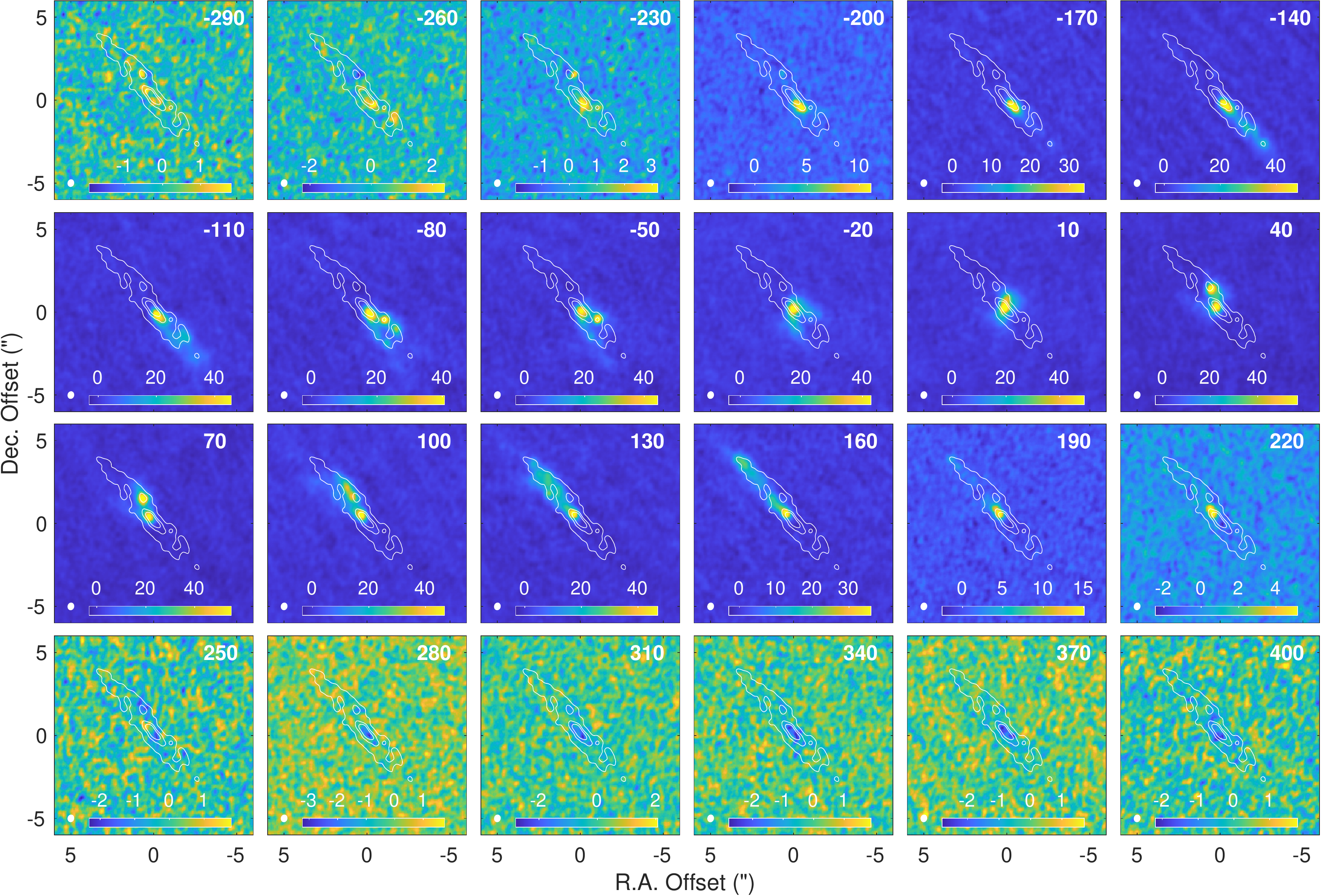}
    \caption{Channel maps for CS \jseven\ emission. Contours, color bars, and beam are as in Figure \ref{fig:channelshcn}. There is no obvious emission associated with any of the outflow features, although this transition is fainter than those of HCN or HCO$^+$.}
    \label{fig:channelscs}
\end{figure*}

\
These considerations strongly suggest that we should apply a factor of two correction to our estimate of the total outflow rate due to outflowing gas that is not easily detectable. \ann{This factor assumes the outflow cone is uniformly populated: in principle the confused half could carry zero mass, or much more mass than the observed half, but in lieu of further information the best approach to correcting for the unobserved material is to assume the cone is uniform.} Therefore the molecular mass outflow rate within $\sim100$~pc of the nuclear region is $\dot{\rm M}_{mol}\approx20$~\msun\,yr$^{-1}$. Given the observed SFR of 4.3~\msun\,yr$^{-1}$ \citep{Bendo2016, Emig2020} the resulting mass loading parameter, the ratio of the mass-loss rate to the star formation rate, is $\eta\sim4.6$. This molecular mass loading parameter is slightly lower but comparable to that measured in NGC\,253 ($\eta\sim8-20$) using a range of CO transitions over distances of 100 to 300~pc from the starburst \citep{Zschaechner2018,Krieger2019}. For M\,82, the mass loading parameter is $\eta\sim8-10$ at a distance of 1~kpc from the midplane, and decreases steadily with increasing distance \citep{Leroy2015}. Therefore, despite the presence of the AGN, the cold outflow in NGC\,4945 appears to be slightly less powerful than what is observed in the two local prototypical starbursts. Nonetheless, the molecular mass outflow rate of $\sim 20$~\msun\,yr$^{-1}$ is much larger than the estimated mass outflow rate for the ionized gas of 1.6~\msun\,yr$^{-1}$ \citep[][Table 5]{Heckman1990}. The CO molecular mass loss originates from a region of $\sim2.3\arcsec$ ($\sim40$~pc) in diameter around the center of NGC\,4945 that shows the brightest 850~$\mu$m continuum identified with the very core of the starburst, although the starburst itself extends beyond that region \citep{Emig2020}. \ann{The fact that the AGN in NGC\,4945 is not energetically dominant \citep{Forbes98,Spoon2000}, and the launching region being extended over an area $\sim40$~pc in diameter, suggest that the molecular outflow is primarily powered by the star formation activity despite the fairly fast speeds measured for a portion of the gas.} 

\ann{Is there a need for momentum or energy input from the AGN to explain the observed outflow, or does the star formation activity suffice? Setting aside the issue of how a molecular outflow of the type we find in NGC\,4945 is driven by star formation, we provide a first answer to this question by comparing the bulk momentum and energy in principle available from star formation. If we take a typical physical outflow speed of $v_{outf}\sim200-300$\,\kmpers, the momentum injection rate necessary to drive 20\,\msun\,yr$^{-1}$ is $\dot{p}_{outf}\sim4,000-6,000$\,\msun\kmpers\,yr$^{-1}$. The net momentum deposition per unit stellar mass formed is $p/m_*\sim3,000$\,\kmpers\ \citep[from the typical momentum deposition per SN and the SN rate of one per 100 \msun\ of stars,][]{Kim2015}, therefore the available momentum injection rate due to the observed SFR$\sim4.3$\,\msunperyr\ is $\dot{p}\sim12,900$\,\msun\kmpers\,yr$^{-1}$, enough to drive the outflow with a coupling  efficiency of $30\%-45\%$. Note that additional sources of momentum (stellar winds, radiation, cosmic ray pressure) are not taken into account in this calculation. The energy deposition rate due to supernovae and stellar winds can be estimated as $\dot{E}\sim2.2\times10^{49}\,(SFR/\msunperyr)$\,erg\,yr$^{-1}$ \citep{Veilleux2005}, yielding a total energy deposition rate for the starburst of $\dot{E}\sim9.5\times10^{49}$\,erg\,yr$^{-1}$. The energy injection rate needed to explain the outflow is $\dot{E}_{outf}\sim(1.6-3.6)\times10^{49}$\,erg\,yr$^{-1}$, which suggests coupling efficiencies of $15\%-40\%$. Therefore neither the momentum nor the energy budgets require AGN input. This analysis, however, does not preclude the AGN from playing a role in the outflow driving.} 

\ann{It has been suggested that gas in molecular outflows can give rise to significant star formation, and even contribute to the formation of stellar halos \citep[e.g.,][]{Gallagher2019}. To first order, star formation would occur in gas where self-gravity plus external pressure overcomes internal thermal and turbulent pressure support plus bulk expansion in the wind background. Under such conditions we would expect the virial parameter, the ratio of twice the kinetic energy to the potential energy of a cloud, to be of order unity \citep{Bertoldi1992}. The virial parameter can be approximated as  $\alpha_{vir}=5\sigma_v^2/(\pi\,G\,R\,\Sigma_{\rm mol})=5\,R\sigma_v^2/(G\,M_{\rm mol})$ \citep[e.g.,][]{Ballesteros-Paredes2020}, where the measured velocity dispersion $\sigma_v$ includes both internal turbulent motions and bulk expansion or shear across the region as the wind expands. We estimate a typical Rayleigh-Jeans brightness of T$_B\sim4$\,K in CO \jthree\ (corresponding to 25 mJy in a beam, Figure \ref{fig:channelsco}) for gas in the outflow, which translates to T$_B\sim8$\,K in CO \jone\ for $r_{31}=0.5$. The observed line widths are large, even at the resolution of 0.26\arcsec\ they tend to be at least $\Delta v\sim100$\,\kmpers\ FWHM in moderately bright regions of region 2, and as low as $\Delta v\sim40-50$\,\kmpers\ in regions 1 and 3 or clumps of region 4. The corresponding velocity dispersion is therefore $\sigma_v=\Delta v/2.35\sim20-40$\,\kmpers. The estimated diameter of structures is $\sim2$ beams, corresponding to $R\sim5$\,pc, and the corresponding column density of \htwo\ using our adopted \xco\ is then N(\htwo)$\sim2\times10^{22}$\,\percmsq, or a surface density $\Sigma_{\rm mol}\sim9.1\times10^{-2}$\,g\,\percmsq\ including the Helium correction (amounting to M$_{\rm mol}\sim3\times10^4$\,\msun). These values then yield $\alpha_{vir}\gtrsim 60$,  much higher than unity. Note that the precise coefficient in $\alpha_{vir}$ depends on the assumed density structure of the clumps \citep{MacLaren1988} and the ellipticity of the clumps \citep{Bertoldi1992} at the $\sim50\%$ level, and there are uncertainties associated with $\Sigma_{\rm mol}$, but the key quantity is the velocity dispersion. It is  possible that there are substructures with significantly smaller $\sigma_v$ that can collapse, or perhaps individual clumps that accrete mass through cooling \citep{Fielding2020,Fielding2021} growing dense enough to collapse. It is also possible that an external pressure P$_e$ can drive the gas into collapse (depending on how quickly the wind is expanding), but that pressure would have to be well in excess of P$_e/k\sim10^6$\,K\,\percmcu\ \citep{Field2011}, or alternatively the clumps would need to be compressed by colliding against a flow obstacle. But for most of the outflow gas it does not seem particularly likely that the observed conditions are conducive to the formation of stars.}

The ultimate fate of the gas is unknown: our observations show outflowing gas only within $\sim100$~pc of the nucleus, but with large velocities suggesting it will reach large distances or even escape. To first order, in a system with a flat rotation curve a particle with $v>\sqrt{2}\,v_{rot}$ has enough energy to escape in a purely ballistic trajectory \citep[i.e., accounting for no decelerations in excess of gravity, which may be present in some systems, e.g.,][]{Martini2018}. The physical velocities for features 1 and 4 implied by our reasoning above ($v\gtrsim350$~\kmpers) and the velocity extent observed for feature 2 ($v\sim450$~\kmpers\ with no corrections, and $\sim600$~\kmpers\ with the minimum projection correction in the cone geometry) are large compared to the rotation of the system 300~pc form the center, $v_{rot}\sim190$\,\kmpers\ \citep{Henkel2018}. 

This reasoning suggests that part of the gas has enough momentum to overcome the gravitational attraction and reach far into the circumgalactic medium of NGC\,4945, although not necessarily in molecular form \citep[e.g.,][]{Leroy2015}. What is clear is that even if this gas is gravitationally trapped and re-accreted by the system it will not necessarily return to the center from which it originated, particularly on a time scale short enough to maintain the central activity. Therefore the duration of the central starburst is determined by the magnitude of this outflow modulo the ability of the barred potential to funnel gas into the central region of the galaxy. The central region of NGC\,4945 contains ${\rm L_{CO}}\simeq27.6\times10^3$\,Jy\,\kmpers,  about $\sim37\%$ of the total CO \jtwo\ luminosity of the galaxy \citep{Leroy2021}. Using a reasonable CO $(2-1)/(1-0)$ line ratio for an active center $r_{21}\sim0.8$, and a CO-to-\htwo\ conversion factor for \jone\ similar to that observed in the starburst of NGC\,253, ${\rm X_{CO}}\approx0.5-1\times10^{20}$\,\xcounits\ \citep[][]{Leroy2015}, this results in ${\rm M}_{mol}\sim3.3-6.5\times10^8$\,\msun. The time scale to remove this much material through the combined outflow and star formation activity is $\tau_{ dep}\sim13.6-26.8$~Myr. For comparison, mid-infrared spectroscopy suggests that the starburst is at least 5~Myr old in NGC\,4945 \citep{Spoon2000}. Note that, with the center containing such a large fraction of the total CO luminosity, gas transport into the central region is very unlikely to drastically change the gas depletion time scale.

\begin{figure}
    \centering
    \includegraphics[width=\columnwidth]{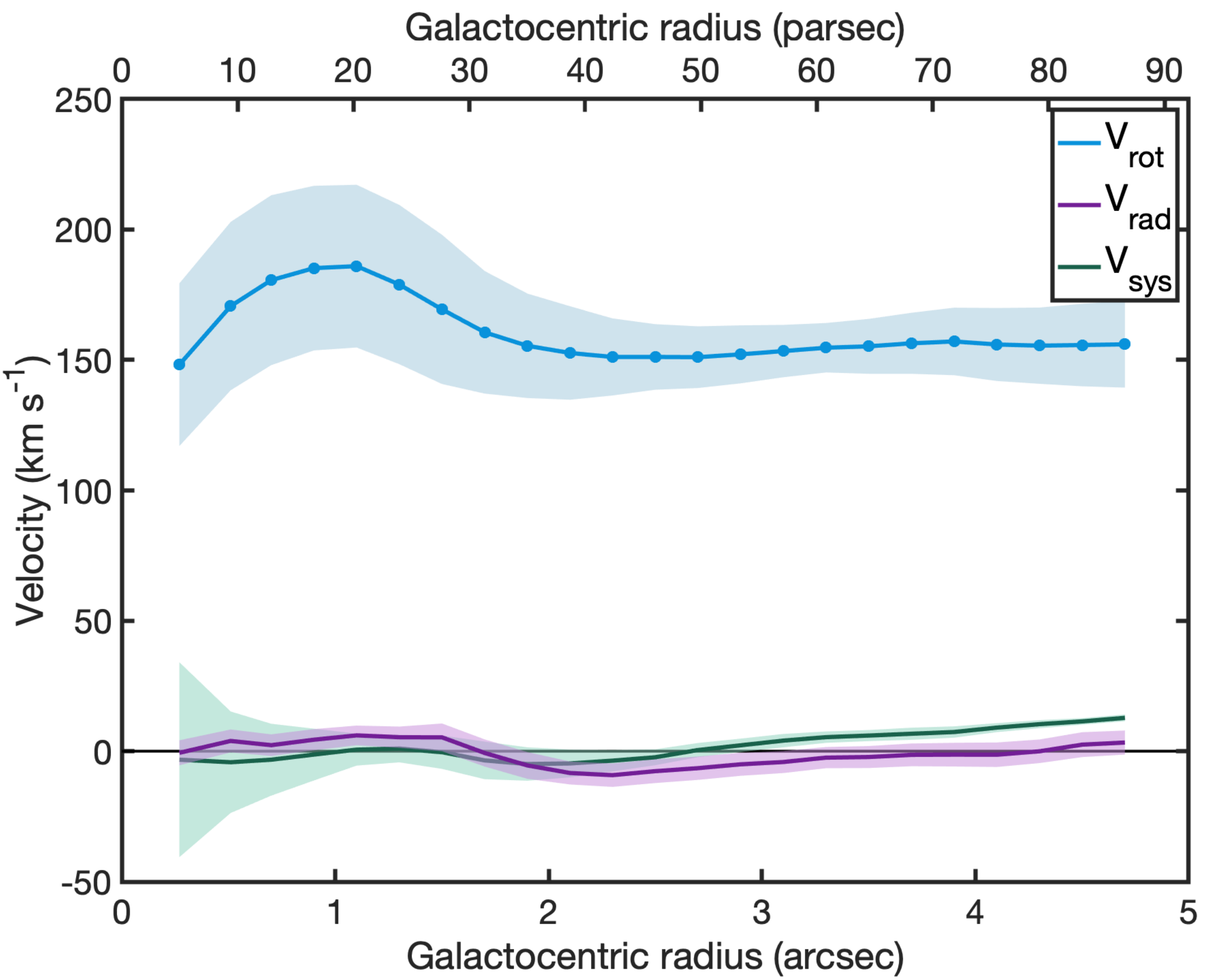}
    \caption{First order kinematic harmonic decomposition obtained from HCN \jfour\ observations. The derived best inclination and PA are $73^\circ\pm3^\circ$ and $225^\circ\pm2^\circ$ respectively. The lines correspond to the rotation, radial, and systemic components of the velocity. The kinematic center is our reference position ($\alpha_{2000}=13^{\rm h}05^{\rm m}27\fs47$ and $\delta_{2000}=-49^{\circ}28{\arcmin}05\farcs6$). The uncertainties shown by the color regions include $\pm0.25\arcsec$ kinematic center positional uncertainty.}
    \label{fig:rotation}
\end{figure}

\subsection{HCO$^+$ and HCN Emission in the Molecular Wind}
\label{sec:HCN}

The molecular wind features are not only visible in CO. The channel maps for HCN and HCO$^+$ \jfour\ display some of the same features despite the fact that the emission in these transitions is considerably weaker (Figures \ref{fig:channelshcn} and \ref{fig:channelshcop}). Over the velocity range for which we have HCN and HCO$^+$ observations we can see the same V-like pattern at $v=-230$ to $-200$~\kmpers, although it is fainter in HCO$^+$ than in HCN, and both are much fainter than in CO. The SE plume of emission is clearly seen at $v=+200$ to $+220$~\kmpers\ in HCO$^+$, although those velocities are missing from our HCN data. 

The CS \jseven, by contrast, shows no hint of wind-related features (Figure \ref{fig:channelscs}). Note, however, that it is fainter than either HCN or HCO$^+$, which may make the features not detectable in our data. 
This situation is also very similar to NGC~253, where the detection of HCN and HCO$^+$ emission in the brightest feature of the NGC~253 wind was reported by \citet{Walter2017} for the \jone\ transitions, with line ratios similar to those measured in the central regions of the starburst.
The channel maps corresponding to the low systemic velocities which contain the bulk of the emission are particularly clean, showing that the effects of confusion and self-absorption are minimal for CS \jseven. 

Are these differences caused by excitation or abundances? From single-dish studies, the abundances for CS and HCN are very similar in NGC~4945, with CS lower by $\sim0.5$~dex \citep[a factor of $\sim3$,][]{Wang2004}. The abundance of HCO$^{+}$ is not well-established, but likely to be \ann{at the order-of-magnitude} level similar to HCN \ann{\citep[note, however, that very high ratios may occur in shocked or irradiated regions;][]{Aalto2015}}. The three high dipole molecules have very similar critical density requirements for their excitation. The critical densities (i.e., the density for which spontaneous de-excitation equals collisional de-excitation) computed for collisions with H$_2$ molecules using the Leiden Atomic and Molecular Database \citep[LAMDA,][]{Schoier2005} in these transitions are very similar: $n_{cr}=3.6\times10^7$\,\percmcu\ (HCN \jfour), $n_{cr}=9.3\times10^6$\,\percmcu\ (HCO$^+$ \jfour),
$n_{cr}=1.5\times10^7$\,\percmcu\ (CS \jseven), all computed at 50~K. Therefore the critical densities for the brighter HCN and HCO$^+$ effectively bracket that of the fainter CS, and density by itself cannot explain the different intensities. The main excitation difference is the kinetic temperature corresponding to the transition upper level, which for CS \jseven\ is $E_{upper}/k\approx65.8$~K versus $E_{upper}/k\approx42.6$~K for either HCN or HCO$^{+}$ \jfour. Therefore we expect the gas emitting in CS to be on average $\sim20$~K warmer than gas emitting at either HCN, HCO$^+$, or for that matter CO ($E_{upper}/k\approx33$~K). In conclusion, the differences in emission seen between CS and HCN or HCO$^{+}$ are likely due to a combination of the somewhat lower abundance of CS and the higher temperature requirements for exciting the \jseven\ transition. Note that it is possible to enhance the emission from high-dipole molecules through collisions with electrons in gas with high electron densities \citep{Goldsmith2017}, but that mechanism would operate for all three of our high dipole molecules.

The detection of HCN and HCO$^+$ in the outflow plumes shows that the gas ejected is chemically complex, and suggests it has high density. \ann{Since the plumes of CO and HCO$^+$ are connected to the disk emission (Figures \ref{fig:channelsco}, \ref{fig:channelshcop}) in position-velocity space, the morphology of the outflow is more reminiscent of molecular gas that has been ejected from the central regions of NGC\,4945 than gas condensed out of a heavily mass-loaded hot outflow \citep{Wang1995,Thompson2016}.
The dynamical time-scale of $t_{dyn}=d^{proj}_{outf}/(v^{proj}_{outf}\,\tan{i_\varphi})\lesssim0.3$~Myr is short, and condensed gas would need to cool from the T$\sim10^6$~K phase and develop a chemistry that contains not only CO but also other molecules such as HCO$^+$ on such timescale. 
%Cloud formation calculations using time dependent chemistry for gas in galaxy disks suggest that the formation of CO takes a few Myr \citep{Clark2012}, significantly lagging the formation of H$_2$. 
The lack of physico-chemical calculations performed for the conditions of condensing gas in an outflow, however, makes it difficult to make absolute statements about the process producing these clouds. 
%Although the conditions \ann{for molecule formation from condensed gas} in a galactic outflow may be conducive to different chemistry, the mismatch between \ann{the time required to produce cool gas from the T$\sim10^6$~K phase plus drive the chemistry needed to create molecules versus the short dynamical timescale} seems difficult to overcome. 
This also impacts our understanding of the physical conditions in the gas.} The abundance of HCO$^+$, for example, could be boosted in slow stellar outflows through the combination of CH with O, but calculations predict slow formation in fast stellar outflows  with velocities more comparable to our galactic outflow \citep{Taylor1995}. Additionally, unless electron collisions dominate the excitation of HCN and HCO$^+$ \citep{Goldsmith2017}, the densities of the molecular gas in the cool wind are high enough to produce significant emission from transitions with $n_{cr}\sim10^7$~\percmcu. In this regard the situation in NGC\,4945 is similar to that in other galaxy outflows with HCN or HCO$^+$ \jone\ emission \citep{Kepley2014,Garcia-Burillo2014,Viti2014,Aalto2015,Walter2017,Barcos-Munoz2018,Cicone2020}, but even more extreme given the higher rotational levels observed, which suggests that a fraction of the gas has densities of order $n\sim10^5$~\percmcu\ and higher.

\subsection{Steep Central Velocity Rise}
\label{sec:kinematics}
Figure \ref{fig:channelshcn} also shows the velocity field of the HCN \jfour\ emission in the bottom right panel. This map was obtained by Gaussian fitting to the spectral line profiles. For highly inclined galaxy disks with asymmetric line profiles this provides a better estimate of the velocity peak and the rotation velocity than producing a traditional first moment, and visual inspection of the spectral fits shows it performs well in this case. We use a first order harmonic decomposition to fit the kinematics, following the algorithms in \citet{Levy2018} and Cooke et al. (submitted). 

The central 10\arcsec\ region has kinematics that are very well represented by a thin rotating disk with a position angle PA$\simeq45^\circ\pm2^\circ$, inclination $i\simeq73^\circ\pm3^\circ$, center $\alpha_{2000}=13^{\rm h}05^{\rm m}27\fs47$ and $\delta_{2000}=-49^{\circ}28{\arcmin}05\farcs6$, and a peak rotation velocity of $v_{rot}\simeq186$\,\kmpers\ reached at a radius $R\simeq1.1\arcsec\simeq20.3$~pc that drops to $156$~\kmpers\ at larger distances from the center (Figure \ref{fig:rotation}). This is consistent with the geometrical assumptions in our calculations ($i=75^\circ$), and agrees very well with the results of previous kinematic modelling of the central regions of this galaxy performed using data with much lower angular resolution \citep[$\theta\sim2.1\arcsec$;][]{Henkel2018}.

This modeling can be used to estimate the mass in the very center of NGC\,4945. The rotation corresponding to the first measured point, at $R=0.27\arcsec\approx5.0$\,pc is $v_{rot}\approx148\pm31$\,\kmpers. In a spherical potential the enclosed mass corresponding to this size and velocity is $M\sim(2.5\pm1.4)\times10^7$\,\msun, about 20 times larger than the supermassive black hole mass of $M_{BH}\sim1.4\times10^6$~\msun\ within the inner 0.3~pc inferred from water maser measurements \citep{Greenhill1997}. 

This excess is probably related to a combination of the uncertainties in the measurement of the rotation of the innermost point and insufficient resolution to probe the relevant black hole scale, although other contributions (for example, alterations in the line profiles related to absorption against the bright central continuum) may also play a role. For reference, the approximate radius of the sphere of influence for a black hole with $M_{BH}\sim1.4\times10^6$~\msun\ in a bulge with velocity dispersion $\sigma_*\sim70$~\kmpers\ \citep[expected for that black hole mass,][]{Ferrarese2006} at the distance of NGC\,4945 is $r_g\sim0.07\arcsec\approx1.2$~pc \citep{Barth2004}, substantially smaller than the innermost rotation point we probe. Therefore we also expect the stellar enclosed mass from a nuclear cluster to possibly dominate our measurement. Applying the equations in \citet{Emig2020} to their measurements of the central 93~GHz source (source 18 in their Table 1) suggests a possible cluster mass of order $M\sim2\times10^6$~\msun, although with very large systematic uncertainties related to modeling assumptions such as stellar age, the poorly constrained synchrotron fraction, and the escape of ionizing photons from the nuclear cluster into nearby regions. To this we should add the contribution of gas to the enclosed mass, which is significant judging from our channel maps (e.g., Figure \ref{fig:channelscs}). Therefore, constraining the black hole mass \ann{would need} further higher resolution observations.

\section{Summary and Conclusions}
\label{sec:conclusions}

We present ALMA observations at a resolution of $\theta\approx0.26\arcsec$ ( $\sim 5$~pc) of CO \jthree , HCO$^+$ \jfour, HCN \jfour, and CS \jseven\ near the nucleus of one of the nearest AGN and starburst galaxies, NGC\,4945. This galaxy is known to host an ionized gas galactic outflow, the approaching lobe of which is visible in H$\alpha$ and other optical transitions \citep{Heckman1990,Venturi2017,Mingozzi2019} as well as soft X-rays \citep{Schurch2002}. The observations presented here reveal the molecular counterpart of this outflow, visible in CO (Figure \ref{fig:outflow}), HCO$^+$, and HCN, for which we estimate a molecular mass ${\rm M_{mol}}\sim1.3\times10^6$\,\msun. The ``foot points'' of the molecular outflow coincide with the southwest and northeast edges of the brightest 850~$\mu$m continuum, and the outflow arises from a region with a diameter of $\sim2.3\arcsec$ ($\sim40$~pc) at the galaxy center. \ann{Together with the fact that the AGN appears to not dominate the energetics of the central region \citep{Forbes98,Spoon2000}, and its contribution is not necessary to explain the outflow momentum or energy, this suggests that the outflow is predominantly driven by star formation.} The detection of wind emission from several molecules is similar to the situation in another nearby starburst, NGC\,253, where the brightest molecular plume is detected not just in CO \jone\ but also in HCN and HCO$^+$ \jone, high dipole molecules that are traditional tracers of high density gas \citep{Walter2017}. This suggests high gas densities in the outflow plumes, particularly in the case of NGC\,4945 where the \jfour\ transitions have critical densities $n_{cr}\sim10^7$\,\percmcu.

The molecular outflow plumes are coincident with the edges of the visible ionized gas outflow, and appear to also outline the receding outflow cone which, placed behind the galaxy disk, is too highly obscured to be detectable in optical or soft X-ray emission. After accounting for the likely projection effects, we find that the physical velocities in the outflowing gas are large, suggesting that the gas will reach far into the circumgalactic medium or even escape the system. The most prominent of these molecular plumes extends to projected velocities of $450$\,\kmpers\ from systemic, almost 3 times the rotation velocity in this region, out to distances of $\sim100$~pc from the central nuclear starburst. Even with the minimum projection correction allowed by the geometry, the gas in this feature reaches a physical velocity $v\gtrsim600$\,\kmpers. These are large velocities, comparable to the velocities measured for the ionized gas in the approaching lobe of the outflow \citep[$v\lesssim-550$\,\kmpers,][]{Heckman1990}. Note that due to limitations in our tuning we do not have full sensitivity to CO-emitting gas at negative velocities, a limitation that we hope to overcome with planned observations.  

We estimate the mass outflow rates  to be between $1.4$ and $3.6$~\msunperyr\ for each of the four plumes observed. The total, directly detected mass outflow rate is $\dot{\rm M}_{mol}\sim10$~\msunperyr\ before correcting for incompleteness. Our observation of blueshifted outflowing gas on the approaching side of the disk and redshifted outflowing gas on the receding side of the disk in both the approaching and receding outflow cones is consistent with the recently ejected gas sharing the rotation velocity of the central regions of the disk, in addition to a radial outflowing component, consistent with observations of the M\,82 outflow \citep[e.g.,][]{Leroy2015}. 

We show that in highly inclined systems such as NGC\,4945, selection effects determine that we only detect molecular outflowing gas against the background of disk emission at particular azimuthal angles. Therefore, there is an important correction factor for unobserved outflowing gas. We express the requirement for detection as a ``velocity separation'' condition (Equation \ref{eq:velsep}), and we use it to estimate the fraction of detectable emission. This results in at least a factor of 2 correction for NGC\,4945. Our best estimate for the total outflow rate at distances of $\sim100$~pc is thus $\dot{\rm M}_{mol}\sim20$~\msunperyr. Given the observed SFR of 4.3~\msunperyr\ \citep{Bendo2016,Emig2020} this results in a mass loading parameter $\eta\sim4.6$ measured within $\sim100$~pc of the nuclear starburst. \ann{We also investigate the stability of the clumps in the outflow using the virial parameter, finding $\alpha_{vir}\gtrsim60$, strongly suggesting that most of the gas in the outflow is unlikely to form stars.}  

Finally, we report the results of kinematic modeling of the gas rotation curve. Our measurements are consistent with those obtained by \citet{Henkel2018} at lower resolution. We use the innermost point in the rotation curve to probe the mass within 5~pc (0.27\arcsec) of the center, and find $M\sim(2.5\pm1.4)\times10^7$~\msun. This is much larger than the black hole mass reported by \citet{Greenhill1997}, but these measurements are on too large a scale to provide good constraints and the discrepancy is most likely due to the mass contribution of stars and gas from a nuclear cluster and nearby regions.

\begin{acknowledgments}

We wish to thank the anonymous referee for their constructive report, which helped improve this manuscript. A.D.B. thanks Nicol\'as Bolatto for his adept use of 3D modeling software to produce the model in Figure \ref{fig:geometry}, and Sebasti\'an Bolatto for his help with Photoshop. This paper makes use of the following ALMA data: ADS/JAO.ALMA\#2018.1.01236S and ADS/JAO.ALMA\#2016.1.01135.S. ALMA is a partnership of ESO (representing its member states), NSF (USA) and NINS (Japan), together with NRC (Canada), MOST and ASIAA (Taiwan), and KASI (Republic of Korea), in cooperation with the Republic of Chile. The Joint ALMA Observatory is operated by ESO, AUI/NRAO and NAOJ. The National Radio Astronomy Observatory is a facility of the National Science Foundation operated under cooperative agreement by Associated Universities, Inc.

A.D.B., R.C.L., and L.L. acknowledge the support from the National Science Foundation through grant NSF-AST2108140. The work of A.K.L. was partially supported by the National Science Foundation under Grants No.1615105, and 1653300. 

\end{acknowledgments}

\bibliographystyle{aasjournal}
\bibliography{references}

\end{document}